\documentclass[twocolumn,trackchanges]{aastex631}

\usepackage{CJK}
\usepackage{xspace}
\usepackage{booktabs}
\usepackage{lmodern}
\usepackage{slantsc}
\usepackage{comment}
\usepackage{xspace}
\usepackage{longtable}

\newcommand{\msun}{M$_\odot$} 
\newcommand{\mwd}{$M_{\mathrm{WD}}$} 

\shorttitle{Q Branch Merger Remnants Within 100 pc}
\shortauthors{Ould Rouis et al.}

\received{December 3, 2025}
\revised{January 23, 2026}
\accepted{February 2, 2026}

\submitjournal{ApJ}

\begin{document}

\begin{CJK*}{UTF8}{gbsn}

\title{White Dwarf Merger Remnants with Cooling Delays on the Q Branch Lack Strong Magnetism} 

\author[0009-0002-6065-3292, sname='Ould Rouis']{Lou~Baya~Ould~Rouis}
\correspondingauthor{Lou~Baya~Ould~Rouis}
\email{lbor@bu.edu}
\affiliation{Department of Astronomy \& Institute for Astrophysical Research Boston University Boston, MA 02215, USA}

\author[0000-0001-5941-2286]{J.~J.~Hermes}
\affiliation{Department of Astronomy \& Institute for Astrophysical Research Boston University Boston, MA 02215, USA}

\author[0000-0001-9632-7347]{Joseph~A.~Guidry}\altaffiliation{NSF Graduate Research Fellow}
\affiliation{Department of Astronomy \& Institute for Astrophysical Research Boston University Boston, MA 02215, USA}

\author[0000-0002-9156-7461]{Sihao~Cheng~(程思浩)}
\affiliation{Institute for Advanced Study, 1 Einstein Dr., Princeton, NJ 08540, USA}

\author[0000-0001-6098-2235]{Mukremin~Kilic}
\affiliation{Homer L. Dodge Department of Physics and Astronomy, University of Oklahoma, 440 W. Brooks St., Norman, OK 73019, USA}

\author[0000-0002-7729-484X]{Olivier~Vincent}
\affiliation{D\' epartement de Physique, Universit\' e\ de Montr\' eal, C.P. 6128, Succ.~Centre-Ville, Montr\' eal, Qu\' ebec, Canada}

\author[0000-0003-2368-345X]{Pierre~Bergeron}
\affiliation{D\' epartement de Physique, Universit\' e\ de Montr\' eal, C.P. 6128, Succ.~Centre-Ville, Montr\' eal, Qu\' ebec, Canada}

\author[0000-0002-9632-1436]{Simon Blouin}
\affiliation{Department of Physics and Astronomy, University of Victoria, Victoria, BC V8W 2Y2, Canada}

\author[0000-0001-7143-0890]{Adam~Moss}
\affiliation{Department of Astronomy, University of Florida, 211 Bryant Space Science Center, Gainesville, FL 32611, USA}

\author[0000-0002-0009-409X]{Isaac D. Lopez}
\affiliation{Department of Physics and Astronomy, Iowa State University, Ames, IA 50011, USA}

\author[0009-0009-9105-7865]{Gracyn~Jewett}
\affiliation{Homer L. Dodge Department of Physics and Astronomy, University of Oklahoma, 440 W. Brooks St., Norman, OK 73019, USA}


\begin{abstract}

A population of anomalous ultra-massive white dwarfs discovered with Gaia, often referred to as the Q branch, show high (multi-Gyr) cooling delays produced by exotic physical mechanisms. They are believed to be the products of stellar mergers, but the exact origin and formation channel remain unclear. 
We obtained a spectroscopically complete, volume-limited sample of the Q branch region within 100\,pc, and found significant differences in atmospheric composition and rotation rates as a function of tangential velocity. 
In particular, we discover that stellar remnants with the longest cooling delays do not show strong magnetism nor detectable short-period rotational variability, as opposed to what is generally believed for double-degenerate mergers. This indicates that either these white dwarfs arise from a formation channel with no strong magnetism induced, or that the magnetism produced from the merger dissipates over the cooling delay timescales.
Our follow-up photometry has also discovered pulsations in the second and third hydrogen-dominated DAQ white dwarfs, one hotter than $15{,}500$\,K, possibly extending the boundaries of the DAV instability strip for white dwarfs with thin hydrogen layers. 

\end{abstract} 



\section{Introduction}

The Q branch refers to an over-density of ultra-massive white dwarfs ($M_{\rm WD} >$ 1.08\,M$_{\odot}$) on the color-magnitude diagram lining up with the carbon/oxygen-core (C/O) white dwarf crystallization sequence \citep{2019Natur.565..202T}. It was first discovered from Gaia, which revolutionized our understanding of white dwarf cooling evolution thanks to large datasets of white dwarfs \citep{2018A&A...616A..10G, 2019MNRAS.482.4570G,2021MNRAS.508.3877G}. Using the precise proper motions and distance measurements provided by Gaia, \citet{2019ApJ...886..100C} provided the first detailed study of the Q branch region by breaking the sample up by their kinematics, finding that roughly 6\% of white dwarfs passing through the Q branch will encounter cooling delays of at least 6\,Gyr that cannot only be explained by the latent heat release from crystallization or by merger delay times. 

Following years of efforts to understand this cooling delay phenomenon \citep{2019A&A...625A..87C, 2020A&A...640L..11B, 2020ApJ...899...46B, 2020ApJ...902...93B, 2020ApJ...902L..44C, 2021ApJ...919L..12C, 2021A&A...649L...7C, 2022MNRAS.512.2972W, 2022MNRAS.511.5984F, 2023MNRAS.520..364F}, the Q branch cooling anomaly is likely best explained as a byproduct of the distillation of neutron-rich species like $^{22}$Ne  \citep{2020A&A...640L..11B, 2021ApJ...911L...5B, 2024Natur.627..286B}. An excess of $^{22}$Ne allows for a phase separation process that creates buoyant crystals which then float up, displacing $^{22}$Ne-rich liquid down. This process liberates gravitational energy and pauses the white dwarf cooling for timescales greater than predicted by theoretical cooling evolution \citep{2021ApJ...911L...5B}.

The required amount of neon for this cooling delay mechanism greatly exceeds solar abundances, several times higher than $X_{^{22}\rm Ne} \simeq$~0.016 \citep{2023ApJ...955L..33S}. Exactly how such objects get produced is still poorly understood, though a fraction of white dwarfs with multi-Gyr cooling delays could result from single-star evolution with high initial metallicities ($X_{^{22}\rm Ne} >$ 0.025) \citep{2025ApJ...990L..47B}. Merger scenarios of helium white dwarf engulfment by main-sequence or red giant stars can result in roughly 0.7\,\msun\, C/O-core white dwarfs with $^{22}$Ne abundances high enough to distill $^{22}$Ne \citep{2026arXiv260100072R}. 

Considering the progenitor scenario is crucial for confirming the physics behind the Q branch. 
Broadly, understanding the precise evolutionary history of white dwarfs is essential to accurately determine total ages, and keep using white dwarfs as cosmic clocks for cosmochronology (e.g., \citealt{2001PASP..113..409F,2025ApJ...983..158P}). Stellar mergers of any type complicate cosmochronology.

Roughly 50\% of massive white dwarfs (M$_{\mathrm{WD}} >$ 0.9\,M$_{\odot}$) are expected to be merger remnants \citep{2012A&A...546A..70T, 2020ApJ...891..160C, 2023MNRAS.518.2341K}. Specifically, of all merger types, double-degenerate (WD+WD) mergers are responsible for about half of massive isolated white dwarfs \citep{2020A&A...636A..31T}. These remnants are best characterized by strong magnetism from an induced dynamo during the merger (e.g., \citealt{2012ApJ...749...25G}), rapid overall rotation from the conserved angular momentum accelerated at short orbital distances in compact binaries (e.g., \citealt{2021ApJ...906...53S}), and fast kinematics typical of older populations from dynamical motions around the Galaxy (e.g., \citealt{2009A&A...501..941H}). 
White dwarfs may also arise from binary mergers at other phases, from the main-sequence to the red-giant phases. 

White dwarf stars with masses greater than roughly 1.05\,M$_{\odot}$ originating from single-star evolution are expected to have oxygen-neon cores (O/Ne) rather than C/O cores \citep{2007A&A...476..893S, 2018MNRAS.480.1547L}. Yet, the pile up of white dwarfs in the Q branch region (M$_{WD} >$ 1.08\,M$_{\odot}$) aligns with the C/O crystallization onset branch \citep{2019Natur.565..202T}, which starts later than the onset of crystallization in O/Ne-core white dwarfs.
Additionally, the physical process of $^{22}$Ne distillation requires C/O-core white dwarfs to produce the strength of cooling delays actually observed in the region \citep{2024Natur.627..286B}. Thus, the population of white dwarfs on the Q branch with high cooling delays likely harbor C/O cores. 
In order to be ultra-massive with a C/O core, the object needs to have been produced through a merger that avoided carbon ignition \citep{2014MNRAS.438...14D}. The location and shape of the Q branch can only be reproduced by including white dwarfs with very thin helium layers, which is also inconsistent with a single star evolution scenario \citep{2024Natur.627..286B}.

Thus, white dwarfs in the Q branch region with high cooling delays should be ultra-massive merger remnants with C/O cores and super-solar metallicities. \citet{2023ApJ...955L..33S} suggest that such objects can be produced by the merger of a C/O-core white dwarf with the He-rich core of a subgiant star, and that the occurrence of this type of merger could explain most if not all of the observed delays on the Q branch. In this scenario, the initial C/O-core white dwarf and subgiant star do not require abnormal initial metallicities. Rather, the merger produces an extra amount of $^{14}$N that gets burned into $^{26}$Mg and leads to the distillation process along with the extra $^{22}$Ne once crystallization begins. Still, models do not predict exactly how much hydrogen and carbon enrich the helium envelope of the remnant following the merger, which would affect the abundance of $^{26}$Mg produced, and therefore how long the cooling delays can be \citep{2023ApJ...955L..33S}. 

In this paper, we are interested in testing the formation pathways of white dwarfs with high cooling delays in the Q branch region in order to eventually inform the mechanisms responsible for the delays. With this 100-pc volume-limited study, we present evidence that ultra-massive white dwarfs with high cooling delays do not show strong magnetism, and make up a distinct population of merger remnants with thin hydrogen layers, weak magnetism, and slow rotation. 

In Section~\ref{sec:subsets}, we detail the sample selection for the Q branch white dwarfs. In Section~\ref{sec:observations}, we describe the spectroscopic observations. In Section~\ref{sec:spec}, we present the spectral type distributions of kinematically selected subsets of white dwarfs in the Q branch region, describing the atmospheric compositions and magnetism of each sample, alongside a comparison sample of canonical-mass white dwarfs within the same temperature range. In Section~\ref{sec:variability}, we investigate the photometric variability of white dwarfs in the Q branch region to constrain the fraction of rapid rotators and pulsators. In Section~\ref{sec:merger_history}, we discuss the merger history channels which cannot lead to the high cooling delays Q branch merger remnants, and propose plausible pathways.


\section{Sample Selection}\label{sec:subsets}

\begin{figure*}[t]
  \centering
  {\includegraphics[width=1\textwidth]{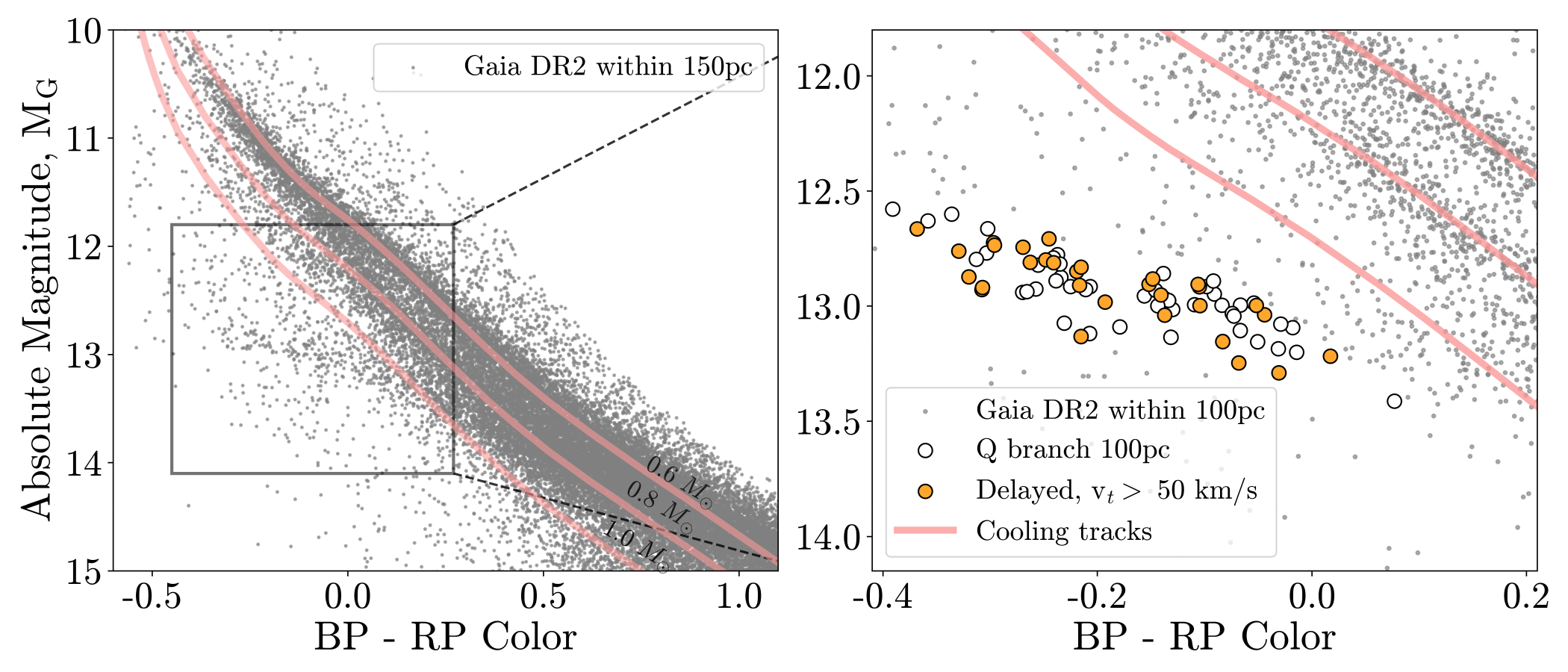}}
  \caption{Color-magnitude diagram of the Gaia white dwarfs (gray points, \citealt{2019MNRAS.482.4570G}) within 150\,pc (left) and 100\,pc (right). The left panel highlights the over-density referred to as the Q branch. The right panel zooms in on the region to show our full sample of 75 white dwarfs in the Q branch region within 100pc as defined in \citep{2019ApJ...886..100C}, where the targets filled in orange make the delayed subset with high cooling delay as indicated by their high kinematics ($v_{\mathrm{t}} >$~50 km\,s$^{-1}$).
  White dwarfs of a given mass are expected to cool from the top left to the bottom right along theoretical cooling tracks shown here (solid pink lines) for 0.6 M$_{\odot}$, 0.8 M$_{\odot}$, and 1.0 M$_{\odot}$ \citep{2020A&A...640L..11B}.} \label{fig:cmd}
\end{figure*}

The region of the color-magnitude diagram we refer to as the Q branch is selected as described in \cite{2019ApJ...886..100C} from Gaia sources. We choose to limit this study to 100\,pc in order to complete a spectroscopic sample.

The boundaries of the Q branch region are defined in \cite{2019ApJ...886..100C}, and were determined using Gaia DR2 parallaxes and magnitudes. If we update all parameters to the most recent Gaia data release (Gaia DR3, \citealt{2021MNRAS.508.3877G}), the boundaries are slightly shifted, and a dozen objects fall out of our sample boundaries, while eight new objects fall inside. For continuity with the observational work on the Q branch, we choose to use Gaia DR2 parameters. The overall sample size and statistics are not significantly affected by adopting the updated parameters.

Our sample of white dwarfs on the Q branch region within 100\,pc contains 75 objects distributed across the entire sky, presented in Table~\ref{tab:sample}. A color-magnitude diagram is presented in Figure~\ref{fig:cmd}. 

\subsection{Kinematics of the Q branch}\label{sec:kinematics}
Ultra-massive white dwarfs should have slow transverse velocities because they are young ($\tau_{\mathrm{total}} <$ 2\,Gyr), based on age-velocity-relations \citep{2009A&A...501..941H}. Abnormally high tangential velocities can be used as a proxy to identify merger products, as the age reset of merger events causes the remnant white dwarf to appear younger than its true total age. 
In the Q branch region, a population of white dwarfs with abnormally high kinematics indicates additional cooling delays, estimated to be at least 6\,Gyr. Such delays cannot be explained only by a merger delay \citep{2019ApJ...886..100C}.

To isolate the white dwarfs with high cooling delays, we divide the Q branch region into two subsets, a kinematically fast population ($v_{\mathrm{t}} >$~50\,km\,s$^{-1}$) that we refer to as the delayed subset and a kinematically slow population ($v_{\mathrm{t}} <$~50 km\,s$^{-1}$) that we refer to as the young subset. Here, $v_{\mathrm{t}}$ is defined as:
\begin{equation}\label{eq}
    v_{\mathrm{t}} =  4.74  \frac{\sqrt{\mu_{\alpha}^2 + \mu_{\delta}^2}}{\varpi}
\end{equation}
where $\mu_{\alpha}$ and $\mu_{\delta}$ are the proper motions in right ascension and declination, and ${\varpi}$ is the parallax. 

The delayed subset is composed of 30 white dwarfs with high cooling delays characterized by fast kinematics. The white dwarfs in the delayed subset are randomly distributed through the Q branch region and highlighted in Figure~\ref{fig:cmd}. The young subset is composed of 45 white dwarfs with the expected velocity distribution of a population with total ages $\tau_{\mathrm{total}} < $ 2\,Gyr, in agreement with theoretical cooling models. The white dwarfs in the young subset are also randomly distributed through the Q branch region and represented by the unfilled circles in Figure~\ref{fig:cmd}.

We opt for a cutoff transverse velocity of $v_{\mathrm{t}} =$~50\,km\,s$^{-1}$ as identified in \citet{2012MNRAS.426..427W} to be a reliable signature of merger remnants. Additionally, 
the kinematics of the class of DAQ white dwarfs (hydrogen-dominated atmospheres with atomic carbon absorption lines) that have been associated with the Q branch cooling delay population \citep{2024ApJ...965..159K} reveal that out of the six published cases within 100\,pc, \cite{2020NatAs...4..663H, 2024ApJ...965..159K, 2025NatAs...9.1347S}, all but one have transverse velocity $v_{\mathrm{t}} >$~51\,km\,s$^{-1}$. The exception, J0551+4135, discovered in \citet{2020NatAs...4..663H}, has a high radial velocity, bringing its total space velocity to 129 $\pm$ 5\,km\,s$^{-1}$ (in the local standard of rest).  

A conservative lower limit of $v_{\mathrm{t}} >$~50\,km\,s$^{-1}$ does not prevent other merger remnants with high radial velocity to contaminate the young Q branch subset, as we only consider transverse velocity in our work.
Therefore, the young subset could include a fraction of O/Ne-core white dwarfs resulting from single-star evolution, along with a contamination of merger remnants: some without noteworthy cooling delays, some with high radial velocity but typical transverse velocity, as well as merger remnants possibly at the very beginning of their time stuck on the Q branch not exhibiting cooling delays just yet. 
Crucially, our kinematic cut ensures that the delayed subset only contains objects with high cooling delays in order to consider this population as independent. 

We compare the simplified transverse velocity measurements ($v_{\mathrm{t}}$) with transverse velocity measurements in Galactic coordinates ($v_{\mathrm{l}}, v_{\mathrm{b}}$) corrected for the motion of the Sun around the Milky Way. All 30 white dwarfs with $v_{\mathrm{t}} >$~50\,km\,s$^{-1}$, have combined $\sqrt{v_{\mathrm{l}}^2 + v_{\mathrm{b}}^2} > $~40\,km\,s$^{-1}$, and 30 out of the 33 objects with combined $\sqrt{v_{\mathrm{l}}^2 + v_{\mathrm{b}}^2} > $~40\,km\,s$^{-1}$ are included in the delayed subset. Choosing either of the two methods does not significantly alter our results, so we adopt $v_{\mathrm{t}}$ parameters for clarity. The transverse velocity values for all white dwarfs in the Q branch region within 100\,pc are reported in Table~\ref{tab:sample}.

\subsection{Comparison Sample}

We also analyze a sample of 94 canonical-mass white dwarfs within 100\,pc to use as comparison. The comparison sample is composed of average mass white dwarfs (0.55\,M$_{\odot} <$ $M_{\mathrm{WD}} <$ 0.80\,M$_{\odot}$) that we select to be in the same temperature range as the Q branch region within 100\,pc assuming photometrically-derived atmospheric parameters with hydrogen-dominated models, or 10${,}$000\,K $< T_{\rm eff} <$ 26${,}$000\,K. We limit the comparison sample to objects in Gaia DR3 with an SDSS spectrum available, and use the SDSS spectral type classifications and photometrically derived atmospheric parameters for H-rich objects from \citep{2021MNRAS.508.3877G}. 
To make sure that the comparative sample only includes white dwarfs compatible with single-star evolution, we limit the kinematics of this sample to the slow subset of the Q branch region, with transverse velocity $v_\mathrm{t} <$~50\,km\,s$^{-1}$. 


\section{Spectroscopic Observations}\label{sec:observations}

We aim to compare the atmospheric compositions of the white dwarfs in each subset of the Q branch region described in Section~\ref{sec:subsets}.

Spectroscopy offers insight on the atmospheric composition of white dwarfs. The strong surface gravity creates stratified layers, where elements heavier than hydrogen or helium sink below the photosphere on timescales shorter than white dwarf cooling ages (e.g., \citealt{1979ApJ...231..826F}). 
With optical spectroscopy, we can identify absorption lines and infer spectral types based on the dominating element present in the atmosphere such as hydrogen (DA), helium (DB), carbon (DQ), or metals such as calcium (DZ) (e.g., \citealt{1983ApJ...269..253S}). When we can identify more than one element in white dwarf spectra, the spectral type reflects the order of the dominating species. 
It is worth noting that from ground-based low resolution spectroscopy in the temperature range of our sample ($T_{\rm eff} >$ 10${,}$000\,K), we are limited in our detection of metals as those transitions become weaker with increasing temperature, so the relative absorption strength decreases (e.g., \citealt{2003ApJ...596..477Z}). 

Spectroscopy also allows us to identify Zeeman splitting of spectral features, which is characteristic of strong magnetism, with typical sensitivity to fields of strength greater than 1\,MG for low-resolution optical spectra \citep{2025ApJ...990...25M}.

We first query the literature for published spectra of our targets. We were able to recover spectra for 53 of the 75 sources in our sample with the Montreal White Dwarf Database (MWDD; \citealt{2017ASPC..509....3D}) published in \citet{2004MNRAS.349.1397C, 2011ApJ...743..138G, 2015MNRAS.454.2787G, 2020ApJ...898...84K, 2023MNRAS.518.3055O, 2024ApJ...974...12J}, as well as archival SDSS spectra \citep{2013ApJS..204....5K}. 

We conduct observing campaigns for the remaining 22 uncharacterized white dwarfs in the 100\,pc Q branch region. 
We also follow up two targets discussed in the literature without public spectra available \citep{1982A&A...108..406K, 2004MNRAS.349.1397C}, and two targets with published spectra but low signal-to-noise ratio in order to confirm their spectral types. 

We include here all 26 new spectra we acquired. Of these, 22 were observed with the Goodman spectrograph on the 4.1-meter Southern Astrophysical Research (SOAR) Telescope with a resolution $R\approx$2000 \citep{2004SPIE.5492..331C}. Three targets were observed with the GMOS spectrograph on the 8.1-meter Gemini South telescope with a resolution R$\approx$1500 \citep{2004PASP..116..425H}. The remaining new spectrum was obtained with the DeVeny spectrograph on the 4.3-meter Lowell Discovery Telescope (LDT) with a resolution R$\approx$1500 \citep{2014SPIE.9147E..2NB}. 
All spectra were reduced and flux calibrated using {\tt PypeIt} routines \citep{2020JOSS....5.2308P, 2020zndo...3743493P}. 
Table~\ref{tab:obs} details the dates, gratings and setups used for all 26 new observations. The new spectra are shown in Figures~\ref{fig:spec_da}, \ref{fig:spec_dq}, \ref{fig:spec_dqa}, and \ref{fig:spec_dh}, sorted by spectral type. 


\section{Spectral types of Q branch white dwarfs}\label{sec:spec}

\begin{figure*}[t]
  \centering
  {\includegraphics[width=1\textwidth]{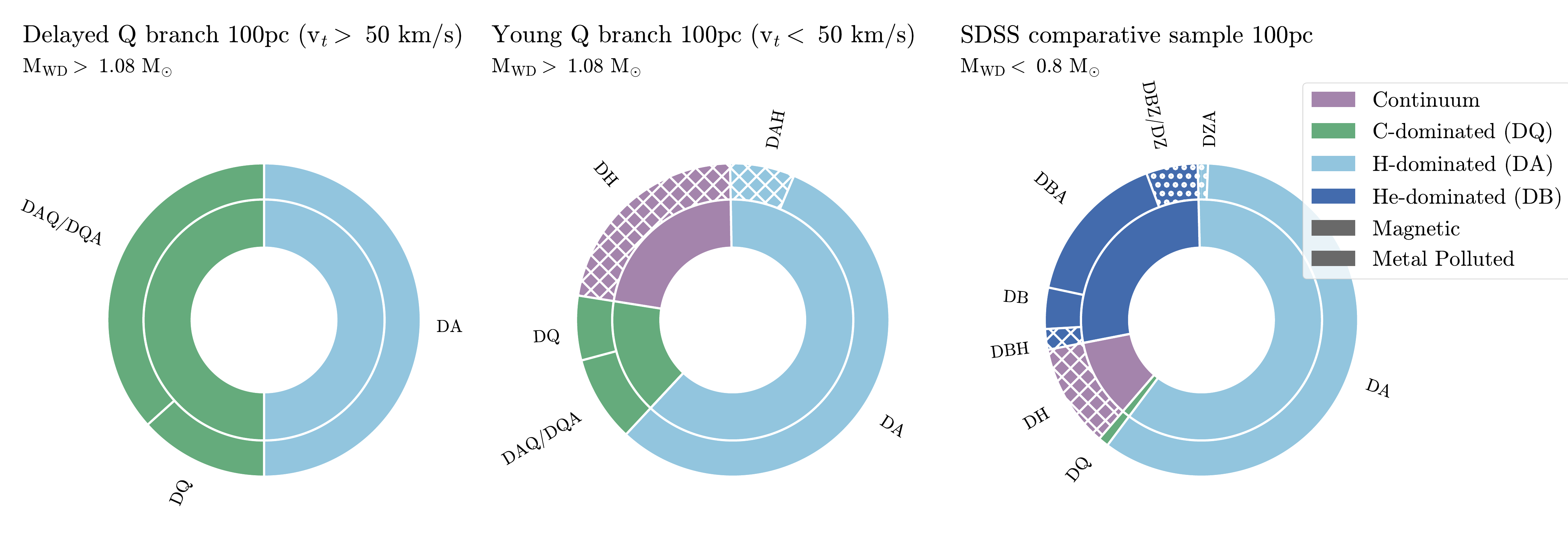}}
  \caption{Sunburst plot of the spectral type distribution for each of the two kinematically selected subsets in the Q branch region and a comparison sample: the delayed Q branch subset ($v_{\mathrm{t}} >$~50\,km\,s$^{-1}$), the young Q branch subset ($v_{\mathrm{t}} <$~50\,km\,s$^{-1}$), and the single star evolution average mass sample from SDSS ($M <$\,0.8\,\msun and $v_{\mathrm{t}} <$\,50\,km\,s$^{-1}$). The inner ring shows the dominant composition from  spectroscopy, and the outer ring specifies the spectral type based on all elements detected in a low-resolution spectra. The physical meaning of each spectral type is described in Section~\ref{sec:spec}. The delayed subset has the highest fraction of white dwarfs with carbon-dominated atmospheres. There are striking differences between the populations in the fraction of magnetism (0\%, 27\%, 14\% respectively), helium-dominated atmospheres (0\%, 0\%, 28\% respectively), and metal pollution (0\%, 0\%, 6\% respectively). } \label{fig:sunburst}
\end{figure*}

In the magnitude-limited spectroscopic sample of white dwarfs in the Q branch region analyzed by \citet{2019ApJ...886..100C}, roughly 50\% of white dwarfs with high cooling delays were classified as DQs (i.e. showing atomic carbon in their atmospheres). From existing spectroscopy in the literature, the authors also noted that most massive DQs are concentrated on the Q branch. The other half of these high-cooling-delay white dwarfs were classified as hydrogen-dominated DA white dwarfs. 

Figure~\ref{fig:sunburst} shows our main classification results, highlighting the atmospheric composition of three different populations, namely the delayed Q branch subset ($v_{\mathrm{t}} >$~50 km\,s$^{-1}$; left panel), the young Q branch subset ($v_{\mathrm{t}} <$~50\,km\,s$^{-1}$; middle panel), and an SDSS average-mass white dwarf sample in the same temperature range for comparison (right panel). All fractional counts are reported in Table~\ref{tab:fractions} for clarity.

\subsection{Delayed subset}

Even with a larger sample than \citet{2019ApJ...886..100C}, we find the same result that 15 out of 30 white dwarfs (50\%) in the delayed Q branch subset show carbon-dominated atmospheres. In most cases we also see hydrogen, characteristic of the DQA and DAQ spectral types (11/15). 

The DAQ subclass is particularly noteworthy. DAQ white dwarfs are expected to have thin hydrogen layer diluted with a deeper carbon-rich semi-convective envelope. 
There have been a total of 16 published white dwarfs with this spectral type, including six DAQs within 100 pc, found in a relatively narrow parameter space \citep{2020NatAs...4..663H, 2024ApJ...965..159K, 2024ApJ...974...12J, 2025NatAs...9.1347S}. These white dwarfs are ultra-massive (0.98\,\msun $<$ \mwd $<$ 1.31\,\msun), hot ($13{,}000$\,K$ < T_{\rm eff} < 23{,}000$\,K), and a large fraction (10/16) have high tangential velocities ($v_{\mathrm{t}} \geq$~50\,km\,s$^{-1}$).
Here, we report a new DAQ white dwarf, WD\,J170145.15$-$524609.22 (shown in Figure~\ref{fig:spec_dq}), bringing the class to seven published objects within 100\,pc, and 17 published DAQs in total. The best-fitting DAQ model has log C/H = 0.43, $T_{\rm eff} = 19{,}104\pm330$\,K and a mass of $1.23\pm0.01$\,\msun, using the photometric fitting method with DAQ model grids as detailed in \citet{2024ApJ...965..159K} and the spectrum fit is presented in Figure~\ref{fig:fit_daq}. This new DAQ becomes the hottest and most massive white dwarf of the class within 100\,pc. 
All DAQ white dwarfs found to date are in the Q branch region \citep{2024ApJ...965..159K}.

The remaining 15 out of the 30 delayed Q branch white dwarfs have hydrogen-dominated atmospheres (50\%), with no other element visible from optical spectra. We note that the DAQ discovered in \citet{2025NatAs...9.1347S} (WD\,0525+526 or WD\,J052950.25+523953.39) is reported as a DA in this study in order to compare the spectral type distribution of the Q branch within the same resolution and wavelength range, and because WD\,J052950.25+523953.39 does not show significant carbon absorption in optical spectroscopy. It is possible that other DAs in the delayed subset exhibit traces of carbon but require more sensitive UV spectra to detect those transitions. Thus, the overall atmospheric composition of the delayed subset is in agreement with the findings from \citet{2019ApJ...886..100C}. 

We do not detect strong magnetism within a limit of 1\,MG among any of the delayed Q branch white dwarfs within 100\,pc (0/30). This suggests that $<$4.5\% of delayed white dwarfs on the Q branch show strong magnetism. 
We investigate archival SDSS spectra for the delayed Q branch beyond 100\,pc, and in this increased sample are able to identify some white dwarfs with strong magnetism, such as the DH WD\,J090632.65+080716.16 which is at a distance of 144\,pc \citep{2021AJ....161..147B}. This reveals that strong magnetism in white dwarfs with high cooling delays is rare. 

\subsection{Young subset}

The young Q branch subset is dominated by hydrogen-rich (DA) white dwarfs. Of the 45 objects in this subset, we observe 25 DA white dwarfs (56\%). Seven white dwarfs are carbon-dominated (18\%), of which more than half also show hydrogen (4/7; spectral type DAQ/DQA). 

We detect Zeeman splitting of hydrogen in three of the 45 young subset white dwarfs (7\% DAH, all fields $B >$ 2.8\,MG). We estimate magnetic field strength based on Zeeman hydrogen splitting diagrams (or ``spaghetti plots'') with the \texttt{Zeeman splitting tool}\footnote{https://github.com/WD-planets/Zeeman\_splitting} \citep{2014ApJS..212...26S, 2023MNRAS.524.4867I}. The estimated field strengths for the three DAH white dwarfs are $B =$~2.8\,MG for WD\,J050709.66+264515.13, $B =$~7.3\,MG for WD\,J133340.35+640627.35, and $B =$~4.5\,MG for WD\,J194736.29$-$310038.84. 

Additionally, 10 white dwarfs have spectra dominated by continua, where very strong magnetism ($\gtrsim$50\,MG) has induced strong non-linear Zeeman effects on the spectrum (22\% DH). We denote these as DH (which are also referred to as MWD for magnetic white dwarfs in \citealt{2021MNRAS.508.3877G}). 
The spectra of cooler white dwarfs ($T_{\mathrm{eff}} < 10{,}000$\,K) often show no absorption lines and are categorized as continuum white dwarfs (DC). But in the Q branch temperature range ($T_{\rm eff} \gtrsim 11{,}000$\,K), we expect hydrogen, helium, or carbon to be visible. Hence, any continuum spectrum in our sample is likely to be a strongly magnetic white dwarf (DH) (e.g., \citealt{2023A&A...670A...2B}).

The overall magnetism fraction of the young Q branch subset is 27\% (12/45, including DAH and DH white dwarfs). While this fraction is remarkably similar to the fraction of magnetic ultra-massive white dwarfs in Solar neighborhood (32\% in \citealt{2024ApJ...974...12J}), this is significantly higher than the field white dwarf magnetic fraction before crystallization, estimated to be $\approx$\,5\% with field strength $B>1$\,MG \citep{2022ApJ...935L..12B, 2025ApJ...990...25M}. 

\subsection{Comparison with canonical-mass white dwarfs}

The third panel in Figure~\ref{fig:sunburst} reveals the spectral type distribution for field white dwarfs within 100\,pc in our stellar neighborhood, the vast majority of which should result from single-star evolution \citep{2020ApJ...891..160C}. Both Q branch subsets show drastic differences from the reference sample. 

The first major difference is the fraction of white dwarfs with helium-dominated atmosphere. The comparison sample shows that 26 out of 94 objects have helium in their atmosphere in this temperature range (roughly $10{,}000 < T_{\rm eff} < 26{,}000$\,K), often with other elements such as hydrogen or heavy metals (28\% spectral types DB, DBA, DBZ, DBAZ). 
But no white dwarfs in either subset of the Q branch show helium. \citet{2024Natur.627..286B} also note the lack of Q branch white dwarfs showing helium-dominated atmospheres. Beyond the Q branch, it has also been observed that massive white dwarfs rarely harbor helium-dominated atmospheres \citep{2024MNRAS.527.8687O,2024ApJ...974...12J, 2025ApJ...979..157K}. 

There is just one carbon-dominated atmosphere white dwarf in the low-mass sample, which is a much lower fraction of DQs than either subset of the Q branch. This is reasonable for a single star evolution subset, as there should be no carbon seen through the photosphere even assuming thin hydrogen and helium layer models \citep{2025arXiv250921717B}, and convection does not explain the presence of DQ white dwarfs for $T_{\rm eff} >$ $10{,}000$\,K (e.g., \citealt{2023MNRAS.525L.112B}).

Magnetism is detected in 13 white dwarfs of the comparison sample, split between continuum white dwarfs (DH, 10/94), and helium-dominated atmosphere white dwarfs with Zeeman splitting (DBH, 3/94). 

The last difference to note is the presence of six white dwarfs with photospheric metals in the comparative sample (6\%, spectral types DZ, DBZ, DZA). Metal pollution is inferred from photospheric transitions of heavy elements (such as calcium, iron, or magnesium) and implies the presence of remnant planetary systems, which are common around average-mass white dwarfs (e.g., \citealt{2014A&A...566A..34K}). We do not detect accreted metals in either subset of the Q branch, possibly tied to the non-detection of helium-dominated white dwarfs on the Q branch since it is easier to detect metals in DB white dwarfs. The implications for planetary survival around Q branch white dwarfs are discussed in Section~\ref{sec:planets}. 

It should be noted that this is not a spectroscopically complete sample, but was selected randomly from serendipitous SDSS spectroscopy, and it should reasonably approximate the underlying population given this reasonably hot temperature range.

\begin{deluxetable}{lccc}[!t]
\tablenum{1}
\tablecaption{Atmospheric Composition of Q branch white dwarfs based on low-resolution optical spectroscopy compared to an average-mass comparison sample in the same temperature range \label{tab:fractions}}
\tabletypesize{\footnotesize}
\tablehead{
    \colhead{Spectral Type} & \colhead{Delayed Q branch} & \colhead{Young Q branch} & \colhead{Comparative}
}
\startdata
DA & 15 & 25 & 57\\
DAH & 0 & 3 & 0\\
DQ & 4 & 3 & 1\\
DQA & 6 & 3 & 0\\
DAQ & 5 & 1 & 0\\
DH & 0 & 10 & 10\\
DB\tablenotemark{{\rm *}} & 0 & 0 & 26\\
\hline
Magnetic & 0 & 12 & 13\\
Metal Polluted & 0 & 0 & 6\\
\hline
\textbf{Total} & 30 & 45 & 94\\
\enddata
\tablenotetext{*}{Also includes DBA, DBAH, DBH, DBZ, and DBAZ spectral types.}
\end{deluxetable}

\section{Photometric Variability on the Q Branch}\label{sec:variability}

Double-degenerate merger remnants are expected to rotate rapidly, having conserved angular momentum from previously being in a compact binary. Theoretical models suggest that WD+WD byproducts should rotate on the order $10-20$ minutes \citep{2021ApJ...906...53S}, much faster than white dwarfs resulting from single-star evolution, which should rotate at periods around 1\,d \citep{2017ApJS..232...23H,2019MNRAS.485.3661F}. We investigate the photometric variability of the objects in our sample as seen by ground-based time-domain surveys to flag any with short-period ($P$ $<$ 2\,hr) variability that could be caused by surface spots that reveal the stellar rotation rate (e.g., \citealt{2024ApJ...967..166S}). We targeted flagged objects with follow-up time-series photometry to verify their variability and also observed objects whose survey photometry is poorly constraining. 

\subsection{Time-Series Photometry}

\subsubsection{Survey Photometry}

We query photometry from the Transiting Exoplanet Survey Satellite (TESS; \citealt{2015JATIS...1a4003R}), the Zwicky Transient Facility (ZTF; \citealt{2019PASP..131a8003M}) and the Asteroid Terrestrial-impact Last Alert System Survey (ATLAS; \citealt{2018PASP..130f4505T}) to search their corresponding Lomb-Scargle periodograms \citep{1976Ap&SS..39..447L,1982ApJ...263..835S} for periodic variability. More precisely, we utilize the PDCSAP flux from the 2-minute and 20-second TESS-SPOC photometry for our search. Our ZTF photometry is sourced from public Data Release 23, which we query on the J2016.0 Gaia DR3-measured centroids using a 3.5\arcsec\ circular aperture. And we use the ATLAS difference images, queried on the same centroids propagated on the Gaia-measured proper motions. 

For each light curve, we compute PSD-normalized Lomb-Scargle periodograms using {\tt astropy} \citep{astropy:2013,astropy:2018,astropy:2022}. This includes partitioning the ZTF and ATLAS photometry by color, treating the $g$, $r$, combined $g+r$, and $c$, $o$, $c+o$ photometry independently. We bootstrap 0.1\% false-alarm probability thresholds with replacement over 10${,}$000 iterations to vet these light curves for periodic variability \citep[c.f.,][]{2018ApJS..236...16V}.

We recover light curves for 74 out of the 75 objects in our full Q branch sample. The only target lacking reliable time-series photometry is WD\,J115618.17$-$675029.94; field crowding ruins the fidelity of its ATLAS images, preventing any reliable analysis. No objects show significant periodic variability in their TESS photometry at periods shorter than 2\,hr. 10 unique objects show periodic significant variability in at least one of their ZTF combined $g$+$r$, $g$, and $r$ light curves. We tabulate these objects and the detected periodicities and amplitudes in Table~\ref{tab:ztf_phot}. We exclude detections of peaks at the daily alias and its respective harmonics.

\subsubsection{Follow-up High-Speed Time-Series Photometry}

We conducted follow-up campaigns that observed 25 rapid-rotator candidates with ground-based, high-speed, time-series photometry. We specifically targeted those objects that showed significant or near-significant variability in their survey photometry and those with poor bootstrapped NOV limits ($\gtrsim$1\%). We observed 18 targets at the 2.1-meter Otto Struve Telescope at McDonald Observatory using the ProEM frame-transfer CCD mounted at Cassegrain focus, another five targets at the 1.8-meter Perkins Telescope Observatory (PTO) using the Perkins Re-Imaging SysteM (PRISM; \citealt{2004AAS...204.1001J}) mounted at Cassegrain focus, and three additional targets at the 4.3-m Lowell Discovery Telescope (LDT) using the Large Monolithic Imager (LMI). The respective plate scales for our observations are 0.36\arcsec\ for ProEM, 0.39\arcsec\ for PRISM, and 0.36\arcsec\ with LMI at 3$\times$3 binning. We include our observing log in Table~\ref{tab:photometry}.
All images were reduced using standard Python-based routines with bias, dark (ProEM only), and flat-field calibration images and the Astropy-affiliated {\tt ccdproc} \citep{2017zndo...1069648C} suite of tools. We performed PSF photometry using {\tt hipercam} \citep{2021MNRAS.507..350D} and used {\tt phot2lc} \citep{2023zndo...8169807V} to extract the optimal light curve. We refer readers to \citet{2025ApJ...992..167G} for a more detailed description of this reduction and light curve extraction process, which we emulate here for each facility.

We similarly period search these follow-up photometry, again computing PSD-normalized Lomb-Scargle periodograms and bootstrapping 0.1\% false-alarm probability significance thresholds with replacement over 10${,}$000 iterations. For the seven objects where periodic variability is detected, we fit sinusoids at those significant frequencies to the time series using {\tt pyriod} \citep{2022ascl.soft07007B}. We report the best-fit frequencies and associated amplitudes in Table~\ref{tab:photometry}. We also report there the bootstrapped significance thresholds for all objects, including those not observed to vary (NOV). We expand on the results of this analysis in the subsequent sections.

\subsection{Rapid Rotators}\label{sec:variability}

We find five rapidly rotating white dwarfs with periods under 2\,hr. 
All five new rapid rotator candidates are part of the young Q branch subset ($v_\mathrm{t} <$~50 km\,s$^{-1}$). Their light curves and associated Lomb-Scargle periodograms are shown in the Appendix (see Figure~\ref{fig:LC}). 

\begin{figure}[t]
  \centering
  {\includegraphics[width=0.47\textwidth]{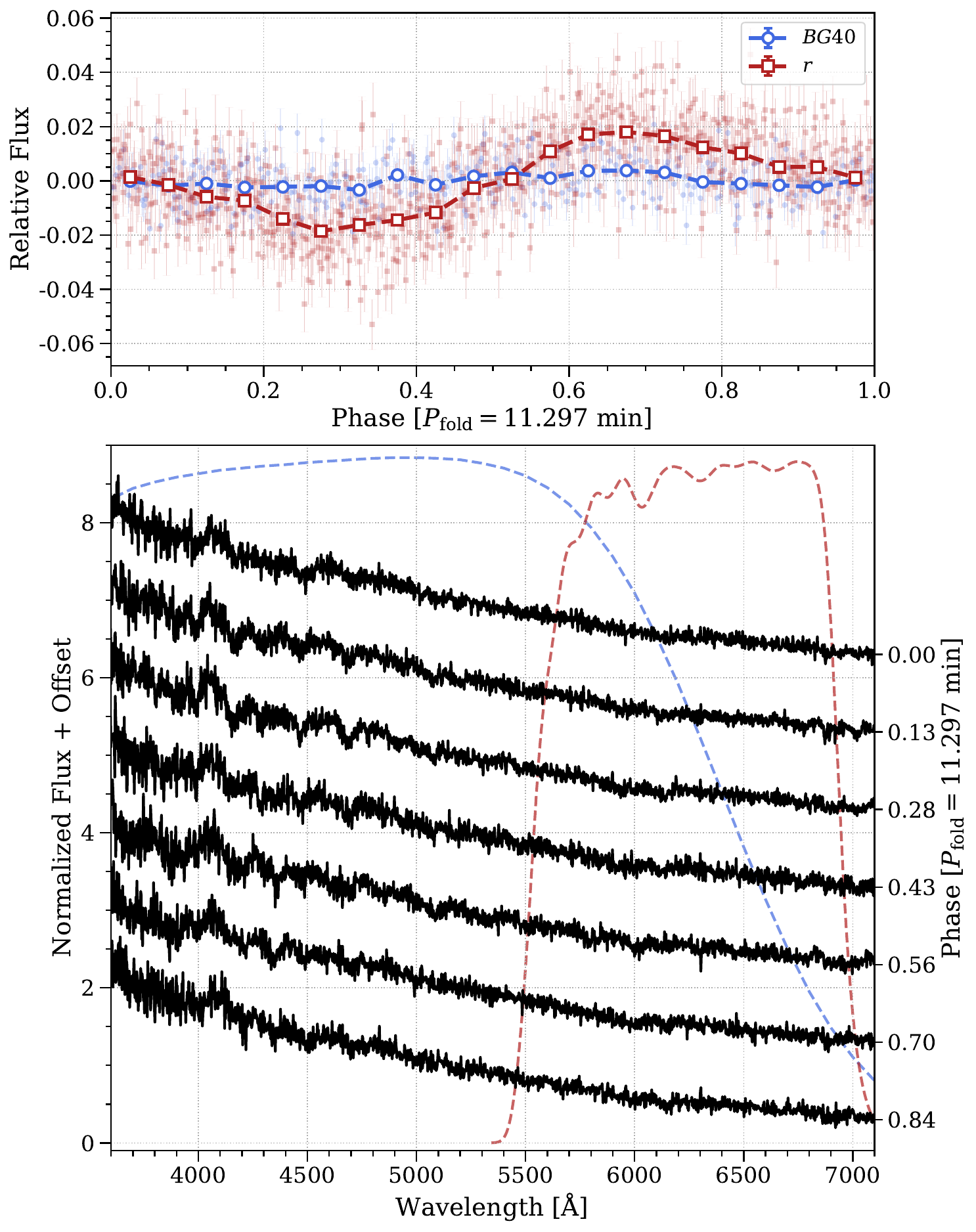}}
  \caption{Comparative analysis of the variability detected for WD\,J145902.72-041157.75. Top: McDonald/ProEM light curves collected on 2025~January~30 ({\em BG40}, blue circles) and 2025~January~31 ($r$, red squares), all folded in phase on the rotation period of 11.297\,min. We overplot binned light curves with respective integration times of 300\,s and 200\,s for the {\em BG40} and $r$ photometry. Bottom: Time-resolved LDT/DeVeny spectroscopy taken on 2025~August~17. We co-add spectra within phase bins of width 0.05. The right-side y-axis ticks track the phase of the 11.297\,min period relative to the start time of the 2025~January~30 light curve. The response curves of the {\em BG40} and $r$-band filters are plotted as dashed  blue and red lines, respectively, both arbitrarily scaled. 
  Our LDT/DeVeny spectra rule out variations at H$\alpha$ driving the 11.297\,min period observed exclusively in the $r$-band, suggesting variations in the continuum flux from strong magnetism.} \label{fig:J1459}
\end{figure}

\begin{figure*}[t]
  \centering
  {\includegraphics[width=1\textwidth]{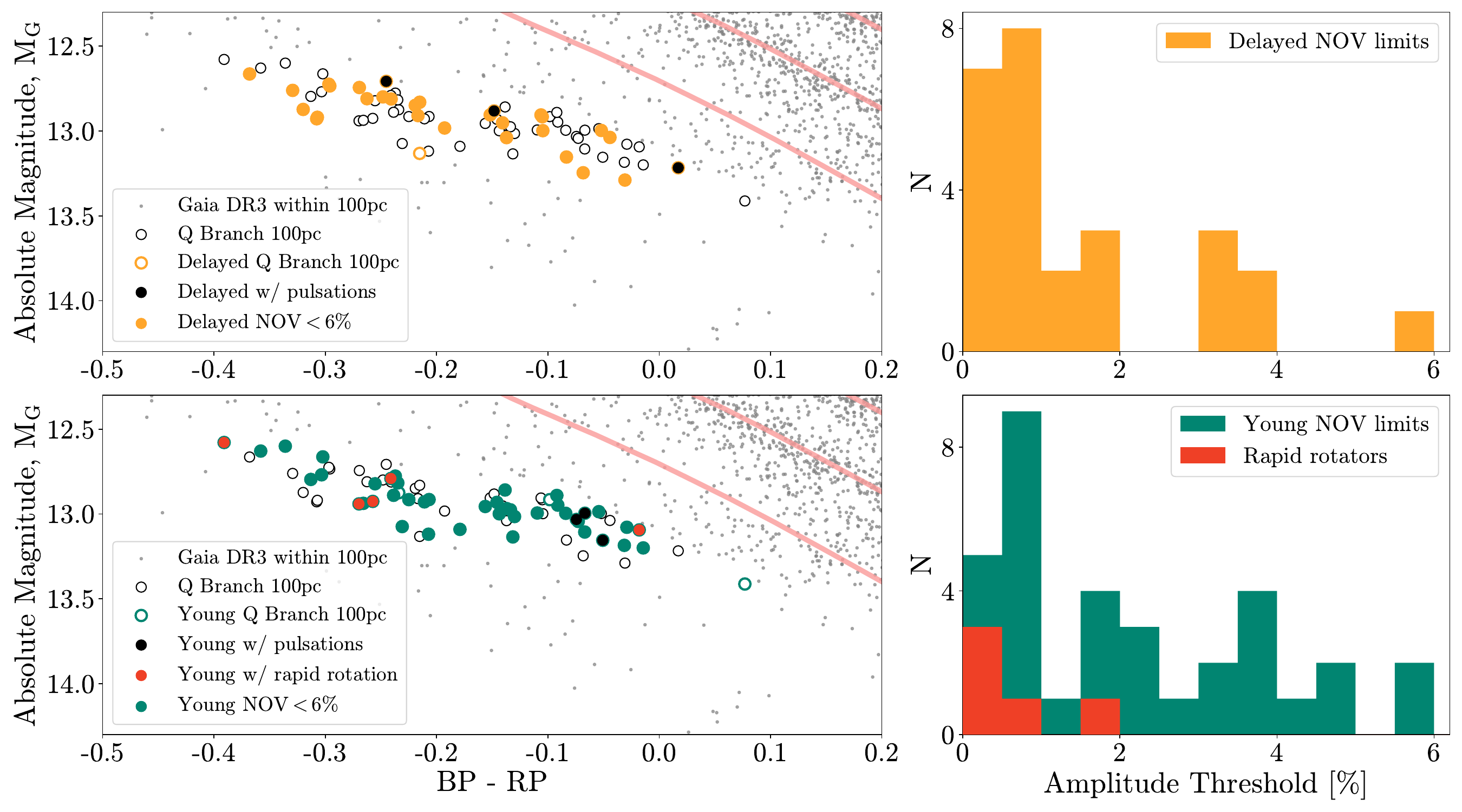}}
  \caption{White dwarfs with rapid rotation and pulsation among the delayed and young Q branch subsets. The left panel shows color magnitude diagrams of the delayed subset (top) and the young subset (bottom) highlighting the targets for which we were able to acquire reliable time-resolved photometry based on bootstrapped NOV limits $<$~6\% (filled targets). The red targets show the spotted white dwarfs with detected periods under 2\,hr, and their observed optical amplitudes are shown in the right panel (red histogram). The black targets show the white dwarfs with pulsations, as described in Section~\ref{sec:pulsating}. The right panel shows the bootstrapped NOV limits for the non-variable objects in the delayed subset (top) and the young subset (bottom). All values are reported Table~\ref{tab:sample}.} \label{fig:photometry}
\end{figure*}

\begin{itemize}
    \item WD\,J002959.00+364834.86 (spectral type DA) shows photometric variability at 65.76\,min (0.44\% amplitude), and no strong magnetism is detected spectroscopically. 
    \item WD\,J062535.30+190244.00 shows photometric variability at 11.05\,min (0.49\% amplitude) and is a strongly magnetic DH. The periodogram shows a signal at what is likely the second harmonic of the 11-min fundamental, indicating a non-sinusoidal light curve. 
    \item WD\,J145902.72$-$041157.75 shows photometric variability at 11.297\,min (1.51\% amplitude) and is a strongly magnetic DH. The periodogram shows a harmonic of the 11-min signal, indicating a non-sinusoidal light curve. This target only shows significant variability on the red (filter SDSS-$r$) and is further explored with time-resolved spectroscopy from LDT/DeVeny as shown in Figure~\ref{fig:J1459}, which suggests that the variations in the continuum flux from strong magnetism cause the variability rather than variations at H$\alpha$. 
    \item WD\,J162157.79+043218.81 shows photometric variability at 36.40\,min (1.0\% amplitude) and is a strongly magnetic DH. Transparency variations complicate the low-frequency region of the periodogram, where there are significant periodicities which could be more fundamental at 1/2 and 2 times this reported period. We inflate our uncertainties on the period for this object given the short dataset. 
    \item WD\,J230844.31+034719.66 (spectral type DQA) has a period of 104.85\,min (0.5\% amplitude), and no strong magnetism is detected spectroscopically. Our 4-hr run covers less than 2.5 cycles, but the periodogram is best explained by a long-period signal and its first harmonic. 
\end{itemize}

While two of the rapid rotator candidates do not show strong magnetism (WD\,J002959.00+364834.86 and WD\,J230844.31+034719.66), these white dwarfs are likely to be weakly magnetic spotted white dwarfs to explain the variability. There are poor constraints on the field strength required for spots to emerge on the surface of the white dwarfs \citep{2015MNRAS.447.1749M, 2018MNRAS.476..933H}.  
Figure~\ref{fig:photometry} shows a color-magnitude diagram for the delayed and young subsets of the Q branch sample which spotlights our photometric variability analysis. We highlight the distribution of rapid-rotator candidates in the young subset along with the distribution of the pulsator candidates in both subsets. We also show the distribution of reliable bootstrapped NOV limits for the targets in both samples that were not observed to vary, put in the context of the amplitudes detected for rapid rotators in the young Q branch subset.

The detection of five rapidly rotating white dwarfs in the young subset of the Q branch is noteworthy (14\% of the young subset). It implies that the objects of the young subset do not all result from single-star evolution. 
We are still limited by observational sensitivity biases, which can be seen from the fact that the typical amplitudes of the rapid rotators in Figure~\ref{fig:photometry} still falls well below many of our NOV limits. Thus, the detection of 14\% rapid rotators in the young subset is a lower limit. Additionally, the relative lack of rapid rotators in the delayed population is notable considering that our NOV limits are roughly equivalent across both samples (the median limit of the delayed population is 0.91\%, compared to 2.01\% for the young subset). 

\subsection{Pulsating White Dwarfs}\label{sec:pulsating}

There are four previously known pulsators among the Q branch sample, which all have notable characteristics. The reported effective temperatures of the following white dwarfs rely on atmospheric modeling detailed in \citet{2024ApJ...974...12J}.
\begin{itemize}
    \item WD\,J013517.57+572249.29 is a $T_{\rm eff} = 12{,}420\pm160$\,K DA with normal kinematics that was reclassified from 7.8\% NOV in \citet{2020AJ....160..252V} as pulsating since it shows multi-periodic oscillations by \citet{2025ApJ...980L...9D}. With 19 pulsation modes, it is the richest pulsating hydrogen-atmosphere ultra-massive white dwarf known to date.
    \item WD\,J055134.45+413529.95 is the first published DAQ, discovered by \citet{2020NatAs...4..663H}, and is included in our young subset, though it has extremely high space motion given its fast radial velocity; it is also likely a merger remnant. 
    \item WD\,J162659.59+253326.79 is a  $T_{\rm eff} = 13{,}200\pm190$\,K DA with normal kinematics that was reclassified from 5.8\% NOV in \citet{2020AJ....160..252V} as pulsating since it shows multi-periodic oscillations by \citet{2025arXiv251009802J}, who also note that this object appears as an outlier compared to expected period-temperature relation for white dwarfs in the DAV instability strip. It is the hottest variable white dwarf in their sample with an exceptionally long period ($1446\pm27$\,s period) (c.f., \citealt{2006ApJ...640..956M}). 
    \item WD\,J165915.38+661032.66 (GD\,518) is a $T_{\rm eff} = 11{,}420\pm110$\,K  DA in the delayed subset which was discovered to pulsate by \citet{2013ApJ...771L...2H} and has fast kinematics ($v_\mathrm{t} =$~59.2 km\,s$^{-1}$), indicating it is likely a merger remnant.
\end{itemize}

At least two of the four previously known pulsators in our sample are consistent with originating from a stellar merger but do not show strong magnetism in their low-resolution spectra. Additionally, observed pulsations imply a low surface magnetic field (less than 50\,kG) as strong magnetism would suppress convection and prevent pulsations from being driven (e.g., \citealt{2024MNRAS.527.6346R}). These merger remnants must originate from an evolution pathway that does not generate strong magnetism.

\begin{figure}[t]
  \centering
  {\includegraphics[width=0.47\textwidth]{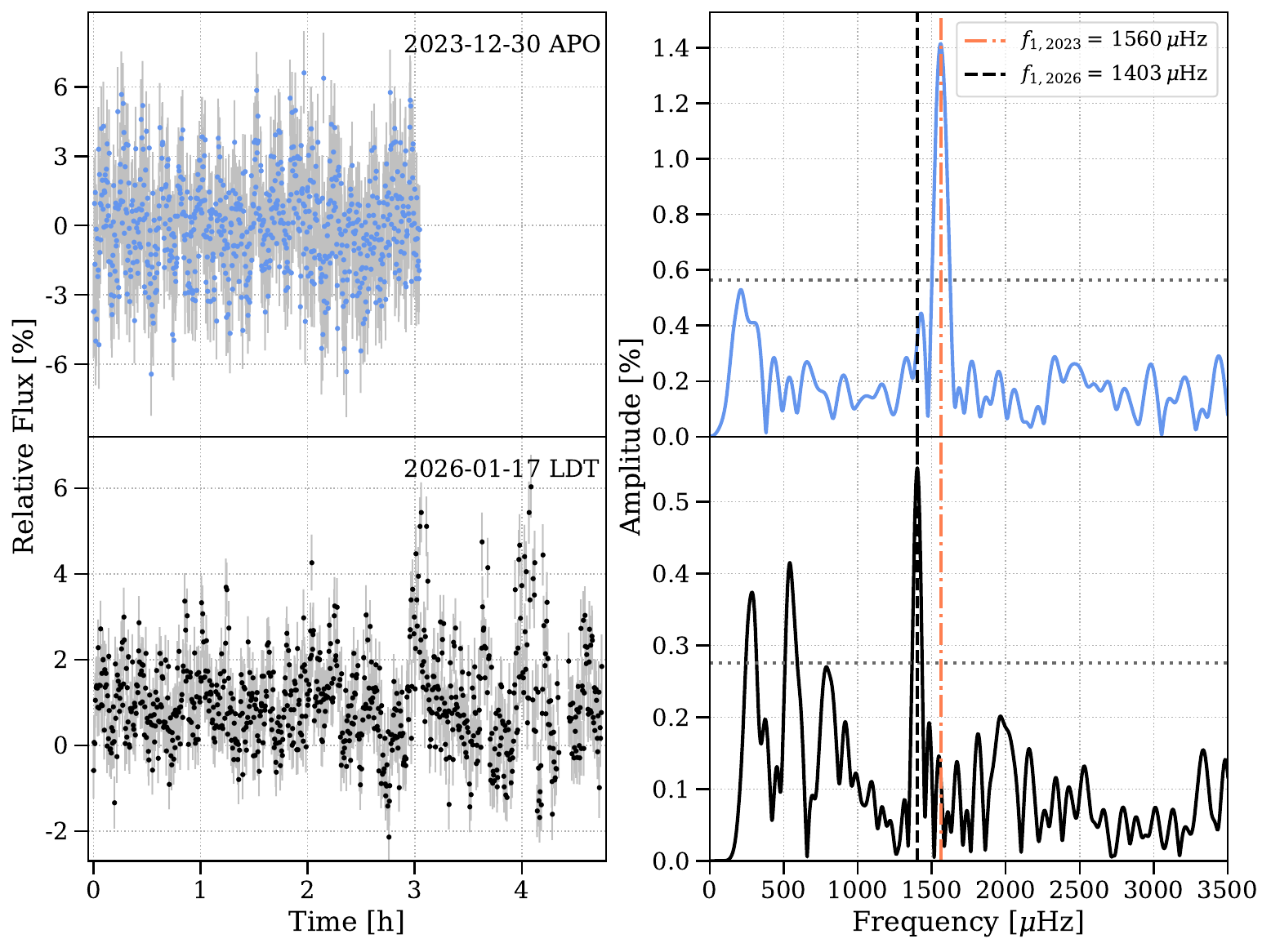}}
  \caption{Variability of the DAQ WD\,J083135.57$-$223133.63 due to pulsations rather than rapid rotation. We show light curves from 2023~December (originally published by \citealt{2024ApJ...965..159K}; {\em BG40}, blue points) and 2026~January ($g$-band, black points) on the left panels. We show the associated Lomb-Scargle periodograms of these light curves on the right panels. The horizontal gray dotted lines delineate bootstrapped 0.1\% false-alarm probability thresholds. The dashed-dotted orange vertical line maps to 1560\,$\mu$Hz, the frequency of the significant peak circa 2023~December, while the black dashed line corresponds to 1403\,$\mu$Hz, the frequency of the significant peak detected in our 2026~January LDT photometry. This period change is inconsistent with a spot and must be due to pulsations, as seen in \citet{2020NatAs...4..663H}. There are no significant frequencies beyond 3500\,$\mu$Hz.}  
  \label{fig:pulsation_0831}
\end{figure}

\begin{figure}[t]
  \centering
  {\includegraphics[width=0.47\textwidth]{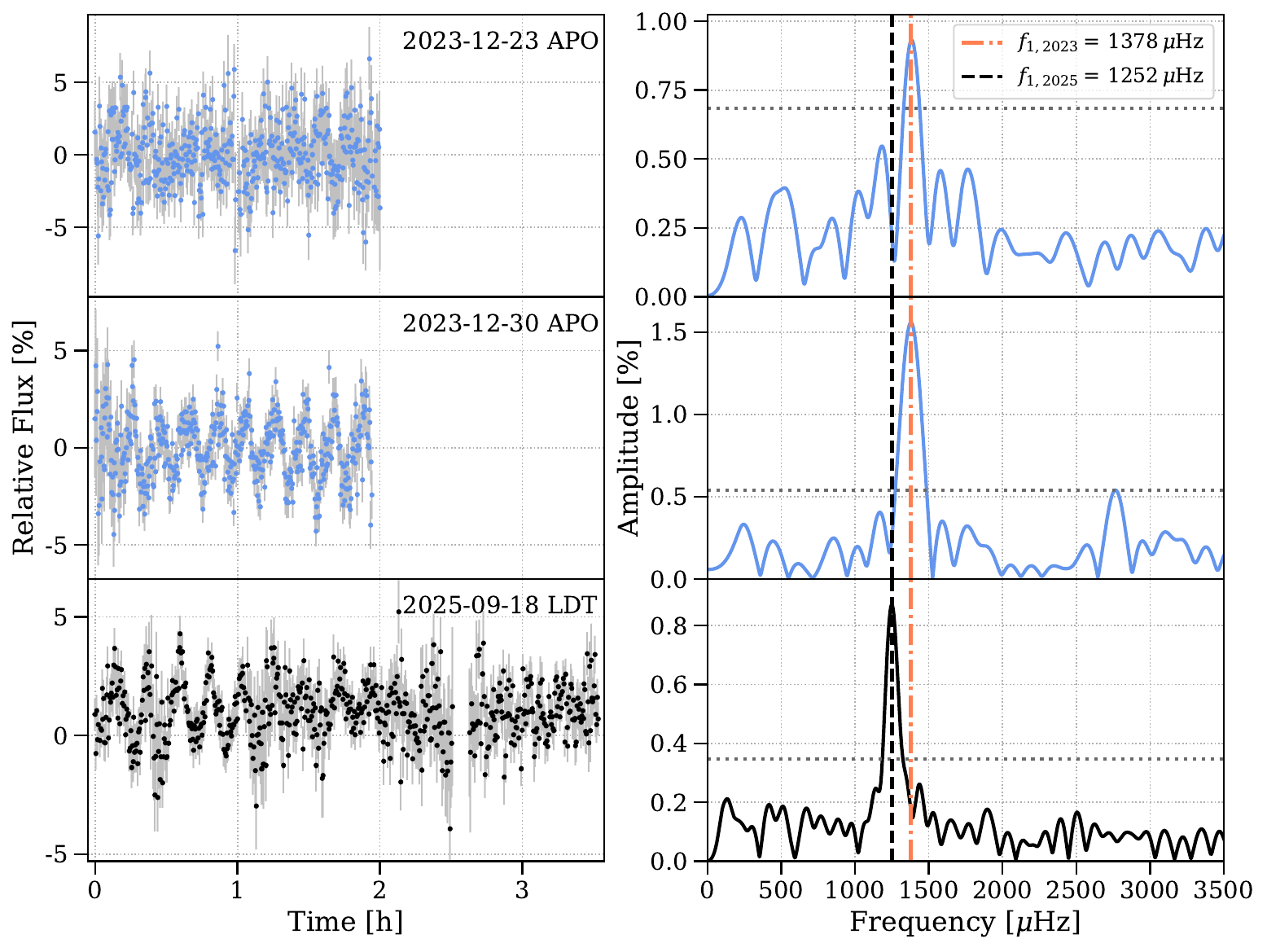}}
  \caption{Variability of the DAQ WD\,J234043.98$-$181945.62 due to pulsations rather than rapid rotation. We plot these light curves and associated Lomb-Scargle periodograms in the same fashion as in Figure~\ref{fig:pulsation_0831}; the blue points are {\em BG40} photometry from 2023~December originally published by \citet{2024ApJ...965..159K}, and the black data represent novel LDT/LMI $g$-band photometry. We observe a stark change in the periodogram, where the dominant mode at 1378\,$\mu$Hz (dashed-dotted orange vertical line) disappears and one appears at 1252\,$\mu$Hz (black dashed vertical line), confirming these are pulsations and not a modulation due to a surface spot. There are no significant frequencies beyond 3500\,$\mu$Hz.}  
  \label{fig:pulsation}
\end{figure}

There are two additional rapid-rotator candidates in the delayed Q branch subset (the DAQ WD\,J083135.57$-$223133.63 and the DAQ  WD\,J234043.98$-$181945.62) identified in \citet{2024ApJ...965..159K} as having photometric variability driven by surface spots. However, our follow-up observations of both targets show periods inconsistent with those previously reported. Observing incoherence in mode amplitudes is characteristic of non-radial pulsations (e.g., \citealt{2017ApJS..232...23H}), so we reclassify WD\,J083135.57$-$223133.63 and WD\,J234043.98$-$181945.62 as pulsating white dwarfs. 
The most recent period is 712.91\,s for WD\,J083135.57$-$223133.63, and 803.46\,s for WD\,J234043.98$-$181945.62. Figure~\ref{fig:pulsation_0831} and Figure~\ref{fig:pulsation} show the period change between the observations from \citet{2024ApJ...965..159K} and our recent follow up from LDT/LMI. In the LDT/LMI follow-up of WD\,J234043.98$-$181945.62, we notably detect two significant frequencies that are closely spaced: $1252.3 \pm 2.8\,\mu$Hz and $1275.5 \pm 4.4\,\mu$Hz (Table~\ref{tab:photometry}), reminiscent of the pulsation spectrum of WD\,J055134.45+413529.95 \citep{2025arXiv251106478U}.

These two new pulsators are part of the delayed subset and classified as a DAQ white dwarf, which means that at least three out of the seven known DAQ within 100\,pc are observed to pulsate.
The remaining four DAQs within 100\,pc are WD\,J020549.70+205707.96 (NOV 0.63\%), WD\,J052950.25+523953.39 (NOV 0.2\%),\\ WD\,J065535.29+293909.66 (NOV 0.41\%), and\\ WD\,J170145.15$-$524609.22 (NOV 3.96\%, from ATLAS photometry). We discuss the implications of these detections for the DAV instability strip in Section~\ref{sec:instability}.


\section{Merger history of Q branch white dwarfs}\label{sec:merger_history}

Our observational analysis of kinematically selected subsets of the Q branch region highlight two distinct populations of ultra-massive white dwarfs. We now connect those differences to merger history pathways. 

\subsection{Delayed Q branch white dwarfs are not consistent with double-degenerate merger remnants}

The white dwarfs in the Q branch region with high cooling delays characterized by fast kinematics ($v_\mathrm{t} >$~50 km\,s$^{-1}$) do not follow our expectation of WD+WD merger remnants. Theoretical models of double-degenerate mergers imply rapid rotation periods and induced strong magnetism \citep{2021ApJ...906...53S, 2012ApJ...749...25G}. Binary population synthesis finds the dominant binary merger type to produce massive white dwarfs ($M_{\mathrm WD} > 0.9$\,M$_{\odot}$) is double white dwarf mergers (about 45\%; \citealt{2020A&A...636A..31T}). We argue that the lack of magnetism in the kinematically delayed anomalous Q branch population could be incompatible with WD+WD mergers. 

\subsubsection{No Strong Magnetism}

Strongly magnetic white dwarfs (field strength greater than 1\,MG) are often found to have masses greater than average white dwarfs \citep{2007AJ....134..741S, 2007ApJ...654..499K}. Differential rotation induced during double-degenerate mergers can produce strong, long-lived magnetic fields, of the order 10$^7$\,G, confined to the outer layers of the merger product \citep{2012ApJ...749...25G}. 
Population synthesis shows that the observed fraction of massive high-field magnetic white dwarfs can be explained by double degenerate stars that merge after common envelope evolution \citep{2015MNRAS.447.1713B}. 

Massive, strongly magnetic white dwarfs tend to have short rotation periods and are consistent with the double degenerate merger channel \citep{2000PASP..112..873W, 2013ApJ...773..136J, 2021Natur.595...39C, 2024MNRAS.528.6056H}. 
Strong magnetism can affect the rotation period by spinning down the star on short timescales which depend on stellar winds during the giant phase in the case of misalignment of rotation and magnetic axes \citep{2012ApJ...749...25G, 2025MNRAS.539.3013C}. What this implies is that strongly magnetic white dwarfs without rapid rotation may still be WD+WD merger remnants. 

Still, none of the 30 delayed white dwarfs on the Q branch within 100\,pc show strong magnetism, to a limit of at least 1\,MG, which suggests that the merger pathways leading to these white dwarfs do not drive strong dynamos.

Finding no evidence of strong magnetism for the delayed white dwarfs with carbon atmospheres (spectral types DQ, DQA, DAQ) contrasts with the large fraction of magnetic hot DQ white dwarfs \citep{2015ASPC..493..547D} which reside above the Q branch (at least 70\% of hot DQs are strongly magnetic; \citealt{2013ASPC..469..167D, 2019ApJ...885...74C}). This could imply that hot DQs evolve into warm DQ in the young subset rather than the delayed subset.
Furthermore, the fraction of magnetism in the young subset of the Q branch subset is consistent with a double-degenerate merger origin.

But strong magnetism in merger products can also be tied to the mass ratio of the progenitors. \citet{2024A&A...691A.179P} shows that unequal-mass merger remnants may produce an extra convection zone in the region that contains most of the magnetic energy, which could destroy the strong, ordered field. They test this phenomenon by simulating double-He-core white dwarfs mergers with a total mass of 0.6\,\msun. The white dwarfs in the delayed subset of the Q branch are ultra-massive, so the specific merger channel in that study could not produce the objects of our sample. Further simulations are required to observe the effects of unequal-mass mergers on double-C/O-core white dwarfs with total mass greater than 1.05\,\msun\ and a C/O-core white dwarf with a He-core white dwarf mergers with total masses up to 1.2\,msun.
Still, in the case where the mechanism of magnetic field suppression applies to other merger channels, the observed lack of strong magnetism could suggest that the Q branch white dwarfs of the delayed subset are produced from unequal-mass mergers.

\subsubsection{No Evidence of Rapid Rotation}

We expect the product of double-degenerate mergers to be rapid rotators. The merger remnant carries the angular momentum from the point of tidal disruption where the binary is at a short orbital period (orbital separations under 1\,hour, e.g., \citealt{1997MNRAS.292..205F, 2020ApJ...894...19R}).
From 1D stellar evolution models of WD+WD mergers, the prediction is for the remnant to reach a characteristic minimum angular momentum after the giant phase, which is then retained assuming conservative evolution. As such, the final products should be white dwarfs with rotation periods of order $10-20$\,min \citep{2021ApJ...906...53S}.

We do not detect rapid rotation via spots for any of the white dwarfs in the delayed kinematic subset of the Q branch region to a median NOV limit of 0.91\%. The typical amplitudes of the modulations observed at rapidly rotating ultra-massive white dwarfs tend to be lower than this value ($A \approx$ 0.7\% from recently discovered rotators in \citealt{2025arXiv251018044W} with the same setup as most of our follow-up observations). 

Using the same sensitivity in all of our photometric study, we were still able to detect five rapid rotators in the young subset. Given our NOV limits and the typical amplitudes of rapid rotators, we expect that the young subset harbor many more rapid rotators that we do not have the sensitivity to detect. When accounting for the median amplitude of the detected rapid rotators in our sample which is 0.5\%, we can compare the rapid rotation detections in both subsets limited to NOV limits under 0.5\%. While this yields extremely small number statistics, it sets a 1-$\sigma$ upper limit of rapid rotation of 14\% rapid rotation in the delayed subset (0/10), compared to a rapid rotation fraction of 50\%$^{+18}_{-15}$ (5/10). This demonstrates that the observed lack of rapid rotation within the delayed population is significant. 
In order to present quantifiable occurrences of the rapid rotation in either subset, better limits are necessary. 

The lack of photometric variability from spots should not be treated independently from the observed lack of strong magnetism, because spots should only be detectable in the presence of weak magnetism, though the exact field strength necessary remains unconstrained \citep{2015MNRAS.447.1749M, 2018MNRAS.476..933H}. 
Yet, our analysis provides further evidence that the white dwarfs in the delayed Q branch subset within 100\,pc are not strongly magnetic, and perhaps inconsistent with WD+WD merger remnants.

\subsubsection{Possible Magnetic Field Decay}

The kinematics of the white dwarfs in the delayed subset of the Q branch reveal older ages than predicted from theoretical cooling tracks. 
As white dwarfs evolve along cooling tracks, magnetism is observed to emerge, resulting in cool older white dwarfs showing a higher fraction of magnetism than young white dwarfs \citep{2022ApJ...935L..12B}. 

The fate of double-degenerate mergers has been modeled on relatively short timescales (e.g. \citealt{2007MNRAS.380..933Y, 2012ApJ...748...35S}). Still, we have poor constraints on the evolution of the field strength as a magnetic white dwarf merger remnant cools. Isolated white dwarfs are expected to see magnetic field decay from ohmic dissipation by a factor of two in 10 Gyr \citep{1973ApJ...184..911F, 1987ApJ...313..284W}. 
Depending on the initial strength of the magnetic field, dissipation can also be driven by Hall drift for fields $10^{12}$\,G\,$ < B < 10^{14}$\,G, or ambipolar diffusion for fields $B \gtrsim 10^{14}$\,G \citep{1998ApJ...506L..61H}. 
Magnetic field decay and thermal cooling effects may impact the thermal characteristics and mass-radius relationship of the remnant \citep{2022ApJ...925..133B}.

Our observations are also compatible with magnetic field decay in double-degenerate products over multi-Gyr timescales, assuming magnetism was induced at the time of merger. The distillation process causing the high cooling delays triggers turbulence within white dwarf cores which may last for billions of years \citep{2024A&A...689A.233L}, and turbulence enhances magnetic diffusivity, which may accelerate the decay of existing magnetic fields \citep{2024ApJ...975...63C}. 
Further modeling is necessary to constrain magnetic decay decoupled from thermal cooling. If magnetic field decay can be shown to be insignificant, that would point to a scenario where white dwarfs with cooling delays in the Q branch arise from a merger channel with no induced magnetism, likely ruling out the double-degenerate merger scenario. 

\subsection{Q branch white dwarfs must be merger remnants}

Even if Q branch white dwarfs with high cooling delays are inconsistent with the double-degenerate merger scenario, a stellar merger must be producing these objects, so other merger histories should be considered.

While cool white dwarfs ($T_{\rm eff} <$ 10,000\,K) with atomic or molecular carbon features can be explained by the dredge-up of carbon from deeper layers after convection kicks in, the same process does not apply to warmer white dwarfs. A different evolutionary pathway is required to produce warm and hot DQs, and growing evidence suggests a merger origin \citep{2015ASPC..493..547D, 2016ApJ...817...27W, 2019ApJ...885...74C, 2023MNRAS.520.6299K}.

Our photometric analysis of the delayed Q branch subset is compatible with the prediction of \citet{2023ApJ...955L..33S} for the cooling anomaly white dwarfs to be the product of a C/O-core white dwarf and the He-rich core of a subgiant, though the expected product of such a merger is relatively unconstrained. Because the orbital distance is wider for a white dwarf and subgiant at the time of merger than for WD+WD binaries, it is reasonable to assume that the angular momentum carried through will not be as high, resulting in longer rotation rates for the merger remnant. This could explain why we do not detect rapid rotation within our sensitivity limits.
The mass ratio of the simulated merger in \citet{2023ApJ...955L..33S} calls for a 1.0\,\msun\ C/O-core white dwarf with a 0.2\,\msun\ He-rich core of a subgiant. The extreme unequal-mass of this merger channel could also help explain the lack of strong magnetism \citep{2024A&A...691A.179P}. 

The high cooling delays observed on the Q branch require super-solar abundances to lead to the distillation process. So far, the only non-isolated-white-dwarf scenario which reproduces the requisite metallicities able to drive long cooling delays is the merger of a massive C/O-core white dwarf with a subgiant. Other merger channels, such as a C/O-core white dwarf and a He-core white dwarf, may create ultra-massive C/O-core remnants in the Q branch region but these do not produce the additional $^{22}$Ne necessary to explain the observed Q branch overdensity \citep{2023ApJ...955L..33S}. 

Our observations alone do not allow us to completely rule out different types of stellar mergers. It is also possible to produce anomalous ultra-massive C/O-core white dwarfs without the need of a merger \citep{2021A&A...646A..30A}. Even though such white dwarfs evolve slower than the predicted O/Ne-core white dwarfs, and therefore appear older on the predicted cooling tracks, they do not recreate the high cooling delays observed in the subset of our study with fast kinematics.

\begin{figure*}[t]
  \centering
  {\includegraphics[width=0.85\textwidth]{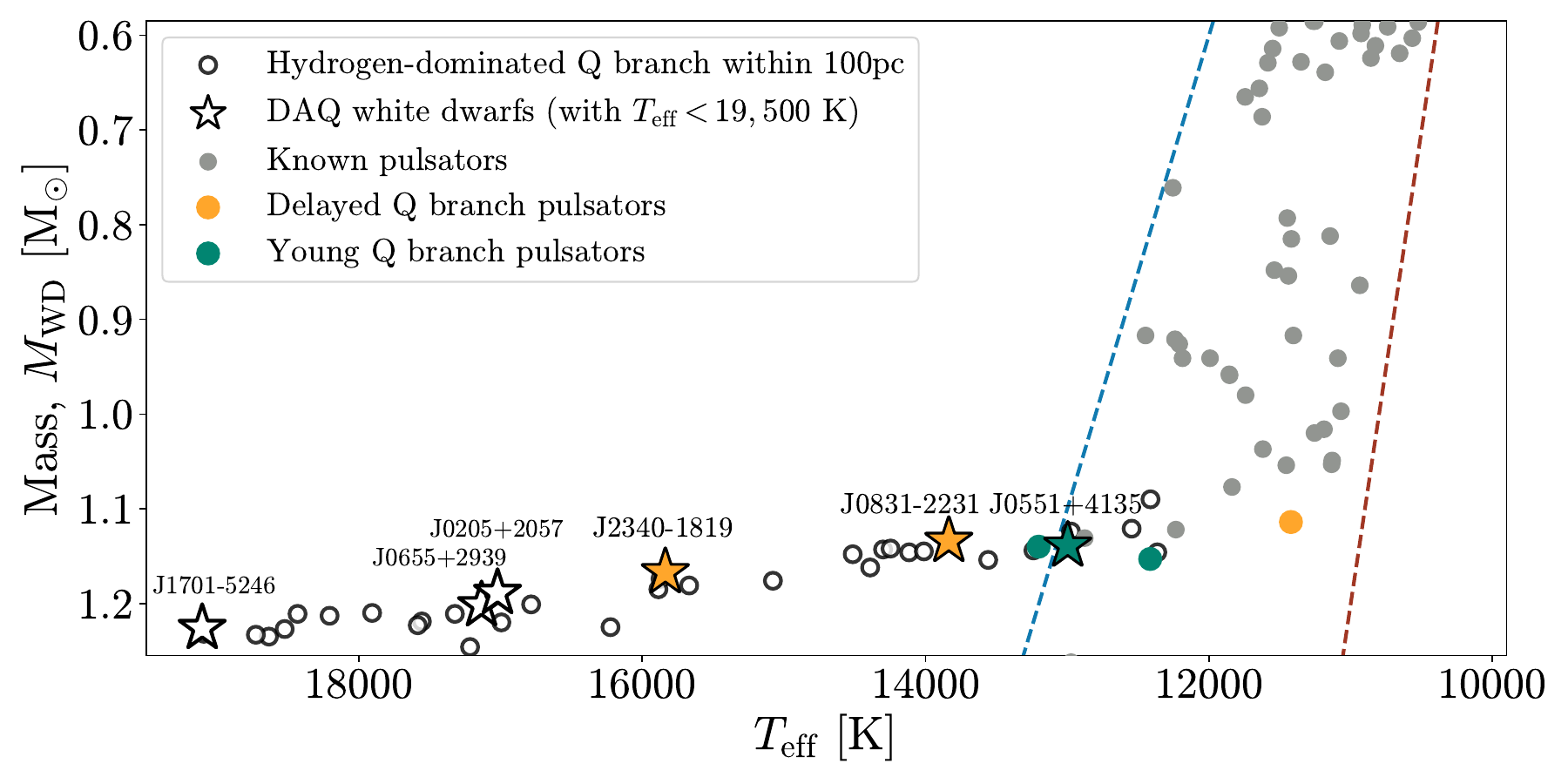}}
  \caption{Known pulsating white dwarfs across the DAV instability strip (gray points) from \citet{2020AJ....160..252V} and \citet{2025arXiv251009802J}. The pulsators of the Q branch are highlighted based on the kinematic subset they belong to (orange for delayed subset, green for young subset). The black circles show the Q branch white dwarfs with hydrogen-dominated atmospheres (spectral types DA, DAQ, DAH). The blue and red dashed lines are the empirical boundaries of the photometric DAV (or ZZ Ceti) instability strip \citep{2020AJ....160..252V}. The stars correspond to the six DAQ white dwarfs with $T_{\rm eff} <$ 19${,}$500\,K within 100\,pc, including our newly discovered target WD\,J170145.15$-$524609.22. Three DAQs within 100\,pc are pulsating.} \label{fig:instabilitystrip}
\end{figure*}

\subsection{No Evidence of Metal Pollution}\label{sec:planets}

There are no detections of photospheric metal pollution from elements like calcium or iron in optical observations around any objects on the Q branch which could imply a lack of remnant planetary systems. 
Although remnant planetary systems around white dwarfs are common \citep{2014A&A...566A..34K}, there is a 4-$\sigma$ decrease in observed metal pollution for massive (\mwd $>$ 0.8\,M$_{\odot}$) white dwarfs \citep{2024ApJ...976..156O, 2025MNRAS.539.2021C}. 
The observed lack of remnant planetary systems on the Q branch is consistent with the previously observed discrepancy in metal pollution as a function of mass, but can also be accounted for only with observational biases. Detecting metals in the atmospheres of DA white dwarfs from lower resolution spectroscopy offers challenges and require higher resolution spectroscopy or UV spectroscopy (e.g., \citealt{2012MNRAS.424..333G}). 

We note that one DA white dwarf in the delayed Q branch subset, WD\,J034703.18$-$180253.49, is believed to be transited by planetary debris on a highly eccentric orbit with a period of $>$\,100\,d \citep{2021ApJ...912..125G}. Higher-resolution spectroscopy in the ultraviolet may reveal photospheric metals on this object, so further investigation is required to fully measure the planetary occurrence on the Q branch.

The Q branch population may also help constrain if the nature of stellar merger events prevents planets from surviving as efficiently as around the isolated progenitors of canonical-mass white dwarfs. Planetary systems can form around binary systems and possibly survive the binary orbital decay within specific configurations \citep{2015PNAS..112.9264M, 2018MNRAS.473.2871V}, though the stability of circumbinary orbits varies across binary mass fraction and binary eccentricity (e.g., \citealt{2021MNRAS.507.4507B}). The dynamical evolution of planetary architecture and the conditions of instability, disruption, and engulfment for planets in interacting binaries is still a topic of active research \citep{2025MNRAS.537..285X}, but the overall conclusion is that tidal forces at play in a merger event create instability for circumbinary planets and increase the chance of escape \citep{2012MNRAS.422.1648V}. 

Higher-resolution spectroscopy is necessary to rule out the presence of heavy elements in the DA white dwarfs of the Q branch region and comment on the remnant planetary system occurrence for ultra-massive merger remnants.

\subsection{Extended Instability Strip}\label{sec:instability}

Our study of pulsating white dwarfs in the Q branch region has revealed a possible extension of the DAV (a.k.a. ZZ Ceti) instability strip. The DAV instability strip refers to a parameter space ($10{,}400$\,K $< T_{\mathrm{eff}} < 12{,}400$\,K for canonical-mass DAs) where H-dominated white dwarfs pulsate with periods ranging from 100\,s to 1400\,s \citep{2019A&ARv..27....7C}. Studies of the massive end of the strip have revealed over 20 pulsators with masses greater than 0.9\,M$_{\odot}$ \citep{2025arXiv251009802J}. 

Many of the pulsating white dwarfs we describe in Section~\ref{sec:pulsating} fall outside of the empirical DAV instability strip boundaries, which are drawn from \citet{2020AJ....160..252V} in Figure~\ref{fig:instabilitystrip}. In particular, we discover pulsations in DAQs that are substantially hotter than the hottest known DAVs, even those of high mass. 

Although DAQ white dwarfs have hydrogen-dominated atmospheres, their hydrogen layers are likely substantially thinner than most DA white dwarfs \citep{2025NatAs...9.1347S}. Thus, it is not necessarily the case that the convective driving of hydrogen (e.g., \citealt{1999ApJ...519..783W}) is responsible for exciting the observed pulsations in DAQ white dwarfs.

For example, some white dwarfs that are accreting solar abundance material in cataclysmic variables may be able to pulsate at higher temperatures, with the driving arising from a mix of hydrogen and helium convection zones \citep{2015AA...575A.125V}. A stability analysis of models of carbon-atmosphere white dwarfs using a nonadiabatic approach found excited $g$-mode pulsations up to periods of 700\,s \citep{2008AA...483L...1F}. The first DAQ, WD\,J055134.45+413529.95, located within the canonical DAV instability strip, is expected to have two convection zones, one superficial zone very near the surface and another much deeper near the region where carbon transitions to hydrogen \citep{2020NatAs...4..663H}.

Driving is unlikely to be very deep: pulsation periods are driven most strongly at periods close to the thermal timescale at the base of the convection zone \citep{1991MNRAS.251..673B}. Since the pulsations observed in DAQs are generally $P<2000$\,s, as in canonical DAVs, it is likely the driving is also happening by the surface convection zone for DAQs. That pulsations are driven in the $T_{\rm eff}>15{,}500$\,K DAQ WD\,J234043.98$-$181945.62 suggests that DAQs have superficial convection at fairly high temperatures. Certainly more theoretical work is needed to understand the DAQ instability strip.

It is worth noting that the DAQs are not the hottest hydrogen-dominated white dwarfs observed to pulsate. That record is held by the so-called hot DAVs near $30{,}000$\,K \citep{2013MNRAS.432.1632K,2020MNRAS.497L..24R}, which are possibly driven by a superadiabatic layer in DAs with thin hydrogen layers that are at the cool end of the DB gap \citep{2007AIPC..948...35S}.


\section{Conclusion}\label{sec:summary}

We conduct a volume-limited spectroscopic and photometric study of white dwarfs in the Q branch region within 100\,pc to investigate an anomalous population of ultra-massive white dwarfs with high (at least 6\,Gyr) cooling delays, first discovered by \citet{2019ApJ...886..100C}. We find significant differences in atmospheric composition and rotation rates as a function of kinematics.

White dwarfs in our delayed subset in the Q branch have fast kinematics ($v_{\mathrm{t}} >$~50\,km\,s$^{-1}$) indicative of high cooling delays, and show carbon-dominated atmospheres more frequently than other populations of white dwarfs, indicating the population typically has very thin hydrogen layers. We expect the ultra-massive ($M_{\rm WD} >$ 1.08\,M$_{\odot}$) delayed white dwarfs on the Q branch to have originated in a merger. However, we do not detect strong magnetism (to a limit $>$1\,MG) nor rapid rotations (to amplitude limits greater than roughly 0.9\%) within the delayed sample. The lack of strong magnetism for the Q branch white dwarfs with the longest cooling delays offers a puzzle regarding the merger progenitor scenario. We hypothesize that these objects may not typically arise from double-degenerate (WD+WD) mergers and instead require a merger channel which does not induce strong magnetism, though we cannot yet rule out that induced strong fields decay over the long cooling delays. Mergers of two white dwarfs with an unequal mass ratio may offer a way \citep{2024A&A...691A.179P} to explain the lack of magnetism in these merger remnants.
Our observations remain consistent with delayed Q branch white dwarfs arising from the merger channel of a C/O-core white dwarf and the He-rich core of a subgiant \citep{2023ApJ...955L..33S}.

The Q branch white dwarfs in our young subset (with slow kinematics, $v_{\mathrm{t}} <$~50\,km\,s$^{-1}$) show a significantly higher fraction of strongly magnetic white dwarfs and white dwarfs detected to have rapid rotation periods, revealing that we can detect such objects in the delayed subset but that such objects are generally rare among white dwarfs with high cooling delays. The young Q branch subset reveals that there are likely WD+WD merger remnants in the Q branch region that may not exhibit extra cooling delays. 

We do not detect any metal pollution within the Q branch region, which is consistent with the lack of massive white dwarfs with photospheric metals indicative of remnant planetary systems (e.g., \citealt{2024ApJ...976..156O}) though further follow up is required as low resolution spectroscopy is insensitive to the detection of heavy metals in DA white dwarfs. 

While the Q branch region falls outside of the DAV instability strip boundaries at much hotter temperatures, we report on six pulsating white dwarfs within our sample, including two DAQ white dwarf (WD\,J083135.57$-$223133.63 and WD\,J234043.98$-$181945.62) whose variability we reclassify from a spot to pulsations in this paper. While the former falls just outside of the DAV instability strip boundaries, the latter is a $T_{\rm eff}>15{,}500$\,K pulsating white dwarf with a hydrogen-dominated atmosphere, thousands of degrees hotter than where we expect convection to be significant enough to drive pulsations for a typical DAV, even at such a high mass \citep{2025arXiv251009802J}. That makes at least three pulsators out of the six published white dwarfs in the DAQ class within 100\,pc, for which we add a seventh in this paper. More work is required to determine the driving mechanism for pulsations in DAQs, which likely harbor very thin hydrogen and helium layers. 

In this work, we have constrained the observational makeup of white dwarf stars in the Q branch of the Gaia color-magnitude diagram within a large enough volume-limited sample in order to better inform the physical origin of the high cooling delays observed in a small subset (6\%) of objects passing through the Q branch.
Our observations remain consistent with the physical picture that the high cooling delays (at least 6\,Gyr) arise from the distillation of neutron-rich species like $^{22}$Ne, as well as a phase separation process that creates buoyant crystals \citep{2020A&A...640L..11B,2021ApJ...911L...5B,2024Natur.627..286B}.
Our results prefer non-double-degenerate merger histories, so we urge future modeling to include other neutron-rich species such as $^{26}$Mg \citep{2023ApJ...955L..33S} to avoid the need for extreme and perhaps unphysical $^{22}$Ne abundances to explain the cooling delays with only neon. Alternative channels for the creation of delayed ultra-massive white dwarfs in the Q branch region remain possible (e.g., \citealt{2023MNRAS.520..364F, 2025ApJ...990L..47B}).

\section{Acknowledgments}

The authors thank Jim Fuller, Evan Bauer, Boris G\"{a}nsicke, Ilaria Caiazzo, and Nicolas Rui for helpful discussions in the preparation of this manuscript. 
L.\,B.\,O.\,R. is supported by the Future Investigators in NASA Earth and Space Science and Technology (FINESST F.5) program under Award No. 80NSSC24K1549.
J.\,A.\,G. is supported by the National Science Foundation Graduate Research Fellowship Program under Grant No. 2234657.
This material is based upon work supported by the National Aeronautics and Space Administration under Grant No. 80NSSC23K1068 issued through the Science Mission Directorate. Support for this work was in part provided by NASA TESS Cycle 7 grant 80NSSC25K7902. 
This work is supported in part by the NSF under grant AST-2508429, and the NASA under grants 80NSSC22K0479, 80NSSC24K0380, and 80NSSC24K0436.

This research is based on observations made with the SOAR telescope associated with AEON program 2025A-866543 proposed through NSF's NOIRLab, as well as observations made on the Lowell Discovery Telescope, Gemini South Telescope, and McDonald 2.1-meter Otto Struve Telescope. The authors thank the telescope operators and observatory staff for their invaluable help and assistance.

The ZTF forced-photometry service was funded under the Heising-Simons Foundation grant \#12540303 (PI: Graham). This work has made use of data from the European Space Agency (ESA) mission Gaia (\url{https://www.cosmos.esa.int/gaia}), processed by the Gaia Data Processing and Analysis Consortium (DPAC, \url{https://www.cosmos.esa.int/web/gaia/dpac/consortium}). Funding for the DPAC
has been provided by national institutions, in particular the institutions participating in the Gaia Multilateral Agreement. This paper includes data collected by the TESS mission. Funding for the TESS mission is provided by the NASA's Science Mission Directorate. 

This research was improved by discussions at the KITP Program ``White Dwarfs as Probes of the Evolution of Planets, Stars, the Milky Way and the Expanding Universe'' supported by National Science Foundation under grant No. NSF PHY-1748958.


\vspace{5mm}
\facilities{ATLAS, Gaia, Gemini:South (GMOS), LDT (DeVeny, LMI), Struve (ProEM), Perkins (PRISM), PO:1.2m (Zwicky Transient Facility), SOAR (Goodman), TESS}

\software{Astropy \citep{astropy:2013, astropy:2018, astropy:2022}, astroquery \citep{2019AJ....157...98G}, ccdproc \citep{2017zndo...1069648C}, hipercam \citep{2021MNRAS.507..350D}, lmfit \citep{newville_2025_16175987}, matplotlib \citep{Hunter:2007}, numpy \citep{harris2020array}, pandas \citep{mckinney-proc-scipy-2010,the_pandas_dev_team_2024_13819579}, phot2lc \citep{2023zndo...8169807V}, PypeIt \citep{2020JOSS....5.2308P}, Pyriod \citep{2022ascl.soft07007B}, scipy \citep{2020SciPy-NMeth}, TOPCAT \citep{2011ascl.soft01010T}}
\clearpage

\appendix

\section{Spectroscopy}
\renewcommand{\thefigure}{A.\arabic{figure}}
\setcounter{figure}{0}

\begin{figure*}[t]
  \centering
  {\includegraphics[width=1.\textwidth]{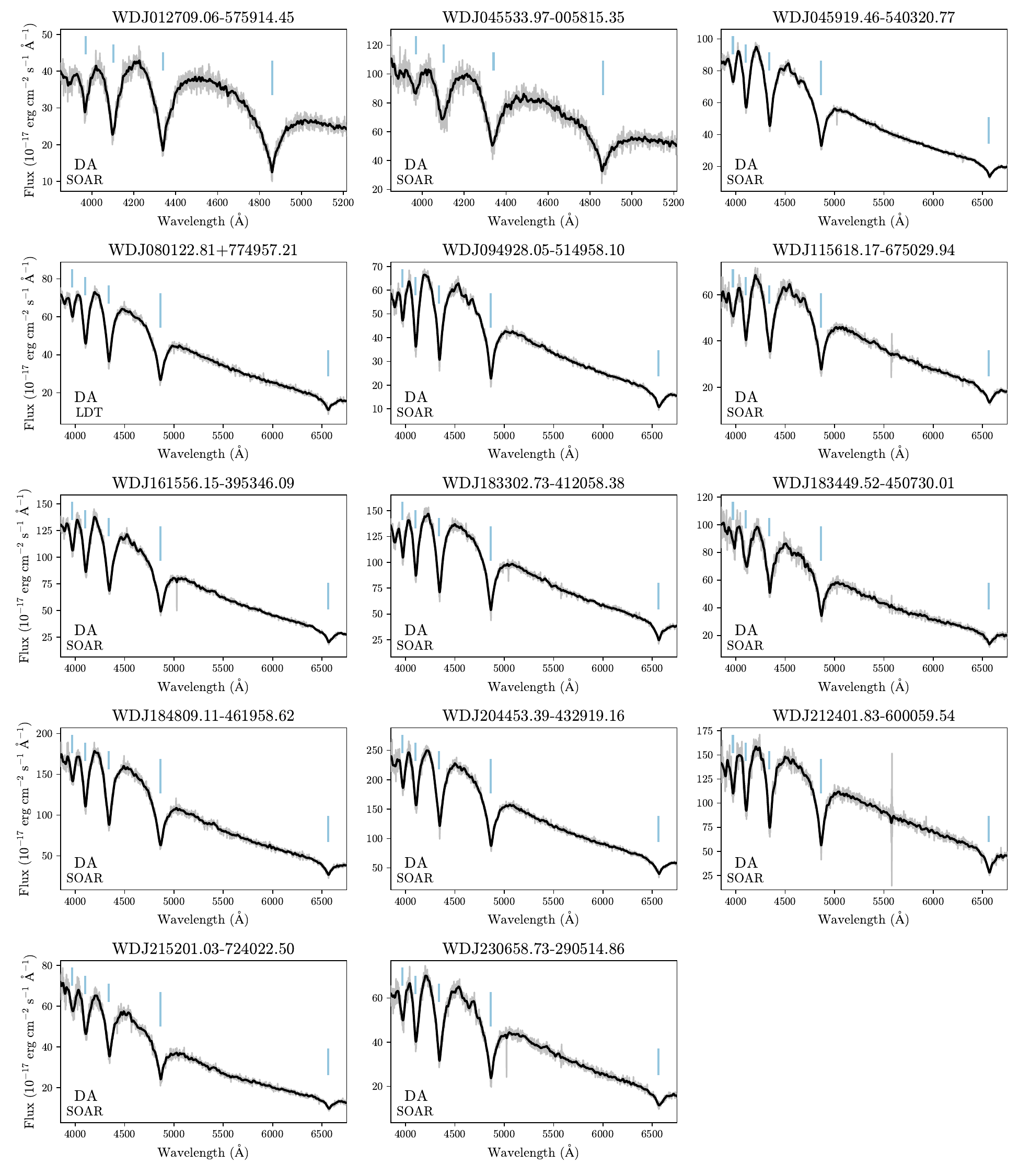}}
  \caption{Spectra from SOAR/Goodman and LDT/DeVeny of 14 new DA white dwarfs, with the Balmer lines indicated by the blue ticks above each spectrum.} \label{fig:spec_da}
\end{figure*}

\begin{figure*}[h!]
  \centering
  {\includegraphics[width=1.\textwidth]{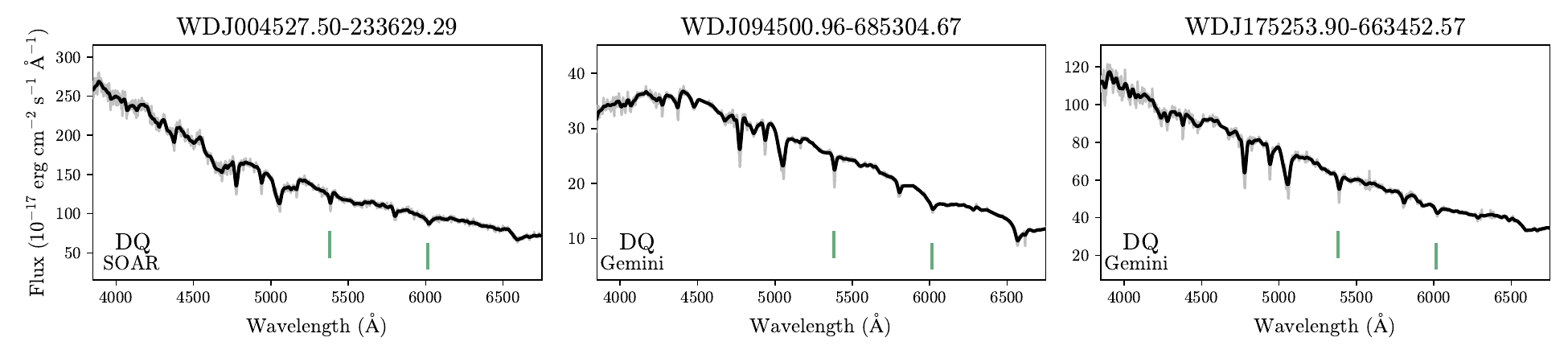}}
  \caption{Spectra from SOAR/Goodman and Gemini/GMOS of three new DQ white dwarfs, with strong C lines indicated by the green ticks below each spectrum.} \label{fig:spec_dq}

\end{figure*}

\begin{figure*}[h!]
  \centering
  {\includegraphics[width=1.\textwidth]{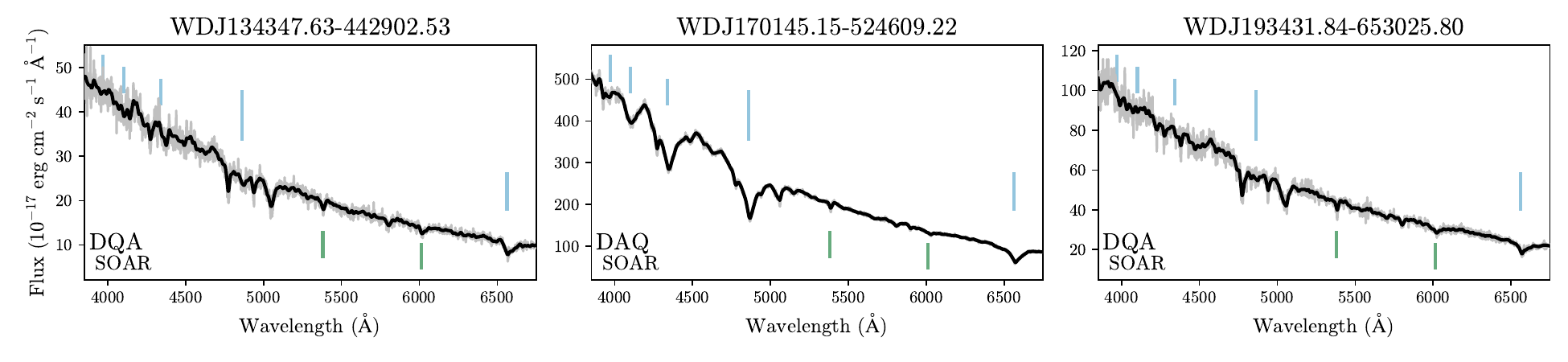}}
  \caption{Spectra from SOAR/Goodman of two new DQA white dwarfs as well as one new DAQ white dwarf, with Balmer lines indicated by the blue ticks above each spectrum and strong C lines indicated by the green ticks below each spectrum.} \label{fig:spec_dqa}

\end{figure*}

\begin{figure*}[h!]
  \centering
  {\includegraphics[width=1.\textwidth]{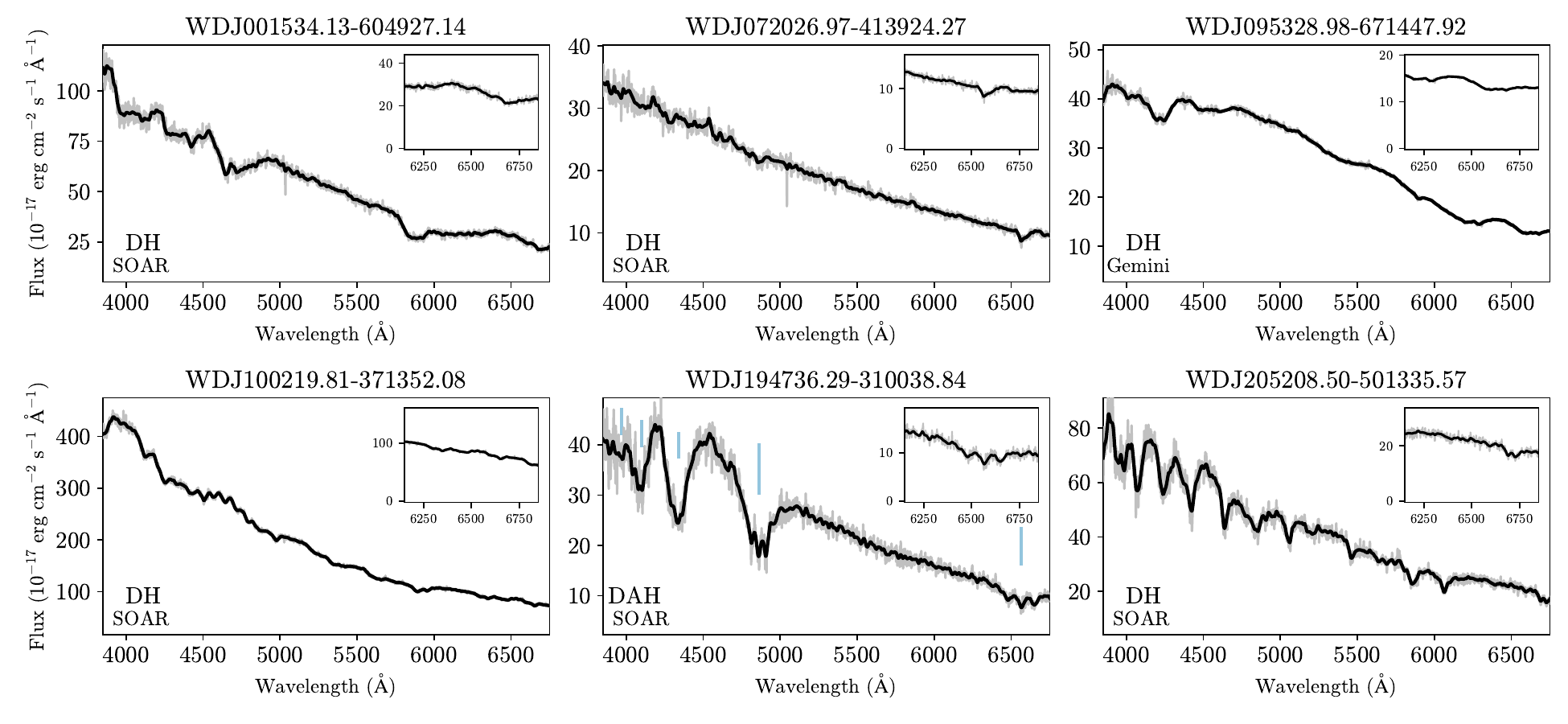}}
  \caption{Spectra from SOAR/Goodman of five new DH white dwarfs as well as one new DAH white dwarf, with Balmer lines indicated by the blue ticks above each spectrum and insets on the H-$\alpha$ region to highlight the Zeeman splitting or the absence of absorption line.} \label{fig:spec_dh}
\end{figure*}

\begin{figure*}[h!]
  \centering
  {\includegraphics[width=1\textwidth]{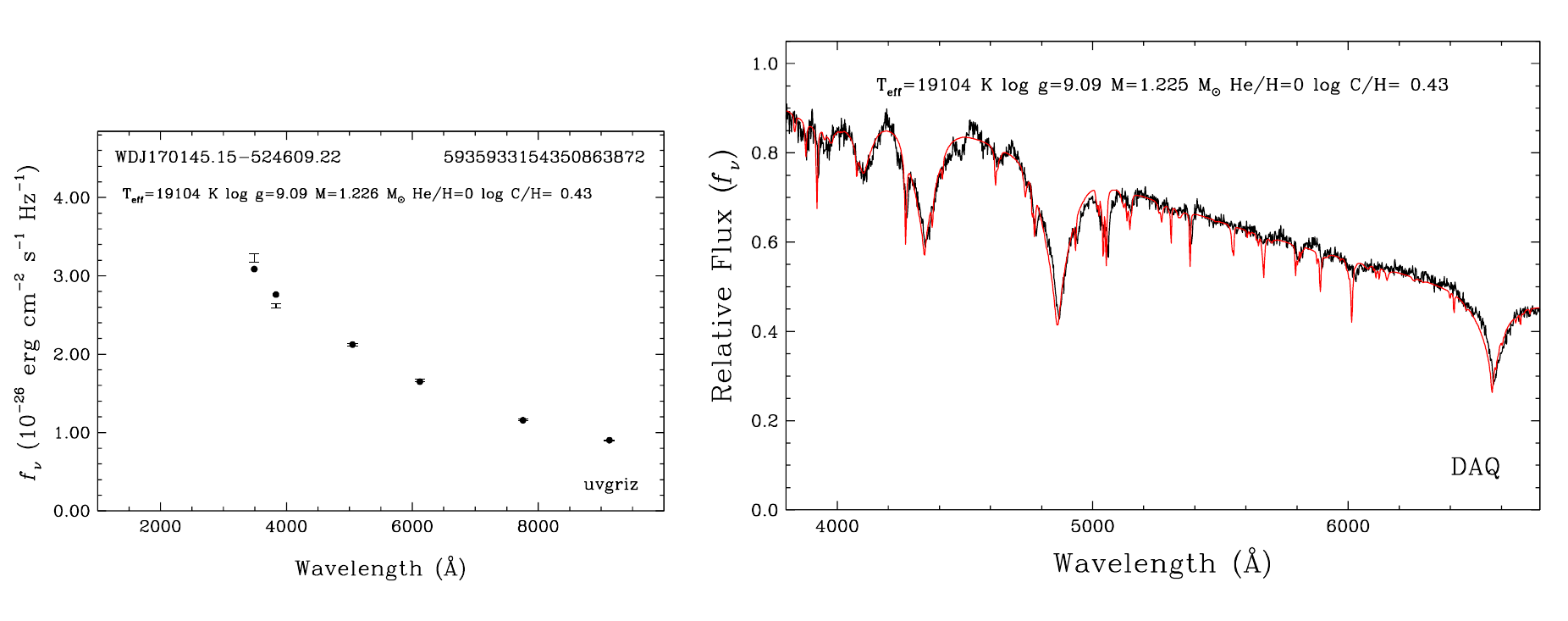}}
  \caption{Model fit of new DAQ WD\,J170145.15-524609.22 with spectral energy distribution from Skymapper photometry (left; \citealt{2007PASA...24....1K}) and SOAR/Goodman spectrum fit (right). The best-fit parameters return log C/H = 0.43, $T_{\rm eff}~=~19{,}104\pm330$\,K and $M_{\rm WD} = 1.23\pm0.01$\,\msun. The DAQ model grids photometric fitting method is detailed in \citet{2024ApJ...965..159K}.} \label{fig:fit_daq}
\end{figure*}

We present all 26 new spectra acquired with SOAR/Goodman, Gemini/GMOS, and LDT/DeVeny. Figure~\ref{fig:spec_da} includes all 14 new DA white dwarfs, Figure~\ref{fig:spec_dq} includes three new DQ white dwarfs, Figure~\ref{fig:spec_dqa} includes two new DQA white dwarfs, as well as one new DAQ white dwarf for which we include a model fit in Figure~\ref{fig:fit_daq}, and Figure~\ref{fig:spec_dh} includes five new strongly magnetic DH white dwarfs, as well as one new DAH white dwarf. The observational details are described in Table~\ref{tab:obs}. 

\begin{deluxetable*}{lccccccc}[htbp!]
\tablenum{A1}
\tablecaption{Observational details of the 26 new spectra we present in this work \label{tab:obs}}
\tablecolumns{8}
\tabletypesize{\footnotesize}
\tablehead{
    \colhead{Target} & \colhead{UT Date} & \colhead{Grating} & \colhead{Slit} & \colhead{Coverage} & \colhead{Exposures (N x $t_{\mathrm{exp}}$) } & \colhead{S/N} & \colhead{Spec. Type}
    }
\startdata
\cutinheadnew{ \textbf{SOAR/Goodman 930-M2}, R $\approx$ 4500} 
WD\,J012709.06$-$575914.45 & 2018-10-27 & 930-M2 & 3'' & 3850-5550 \AA & 7 $\times$ 300\,s & 8.7 & DA\\
WD\,J045533.97$-$005815.35 &  2019-12-06 & 930-M2 & 3'' & 3850-5550 \AA & 7 $\times$ 300\,s & 8.9 & DA\\
\\
\hline
\cutinheadnew{ \textbf{LDT/DeVeny DV2}, R $\approx$ 1500}
WD\,J080122.81+774957.21 & 2020-01-19 & DV2 300g/mm & 1'' & 3000-7400\AA &  5 $\times$ 300\,s & 10.5 & DA\\
\\
\hline
\cutinheadnew{ \textbf{Gemini/GMOS-S B480}, R $\approx$ 1500}
WD\,J094500.96$-$685304.67 & 2024-01-23 & B480 & 1'' & 3600-7000 \AA & 2 $\times$ 525\,s & 21.1 & DQ\\
WD\,J095328.98$-$671447.92 & 2024-01-23 & B480 & 1'' & 3600-7000 \AA & 4  $\times$ 300\,s & 22.3 & DH\\
WD\,J175253.90$-$663452.57 & 2024-01-28 & B480 & 1'' & 3600-7000 \AA & 4  $\times$ 110\,s & 15.8 & DQ\\
\\
\hline
\cutinheadnew{ \textbf{SOAR/Goodman 400-M1}, R $\approx$ 2000 }
WD\,J001534.13$-$604927.14 &  2025-06-05 & 400-M1 & 1'' & 3800-7000 \AA & 4 $\times$ 300\,s & 8.9 & DH\\
WD\,J004527.50$-$233629.29  & 2025-06-08 & 400-M1 & 1'' & 3800-7000 \AA & 4 $\times$ 300\,s & 8.5 & DQ\\
WD\,J045919.46$-$540320.77 & 2025-02-23 & 400-M1 & 1'' & 3800-7000 \AA & 4 $\times$ 480\,s & 9.9 & DA\\
WD\,J072026.97$-$413924.27 & 2025-02-18 & 400-M1 & 1'' & 3800-7000 \AA & 4 $\times$ 480\,s & 14.0 & DH\\
WD\,J094928.05$-$514958.10 & 2025-02-05 & 400-M1 & 1'' & 3800-7000 \AA & 4 $\times$ 480\,s & 10.4 & DA\\
WD\,J100219.81$-$371352.08 & 2025-02-05 & 400-M1 & 1'' & 3800-7000 \AA & 3 $\times$ 300\,s & 16.8 & DH\\
WD\,J115618.17$-$675029.94 & 2025-02-09 & 400-M1 & 1'' & 3800-7000 \AA & 4 $\times$ 480\,s & 11.8 & DA\\
WD\,J134347.63$-$442902.53 & 2025-02-18 & 400-M1 & 1'' & 3800-7000 \AA & 4 $\times$ 480\,s & 9.0 & DQA\\
WD\,J161556.15$-$395346.09 & 2025-02-24 & 400-M1 & 1'' & 3800-7000 \AA & 4 $\times$ 300\,s & 10.2 & DA\\
WD\,J170145.15$-$524609.22 & 2025-02-24 & 400-M1 & 1'' & 3800-7000 \AA & 3 $\times$ 300\,s & 9.7 & DAQ\\
WD\,J183302.73$-$412058.38 & 2025-04-01 & 400-M1 & 1'' & 3800-7000 \AA & 4 $\times$ 300\,s & 11.4 & DA\\
WD\,J183449.52$-$450730.01 & 2025-04-14 & 400-M1 & 1'' & 3800-7000 \AA & 4 $\times$ 480\,s & 10.4 & DA\\
WD\,J184809.11$-$461958.62 & 2025-03-27 & 400-M1$^*$ & 1'' & 3800-7000 \AA & 4 $\times$ 480\,s & 9.9 & DA\\
WD\,J193431.84$-$653025.80 & 2025-04-12 & 400-M1 & 1'' & 3800-7000 \AA & 4 $\times$ 480\,s & 8.7 & DQA\\
WD\,J194736.29$-$310038.84 & 2025-03-21 & 400-M1 & 1'' & 3800-7000 \AA & 4 $\times$ 300\,s & 7.2 & DAH\\
WD\,J204453.39$-$432919.16 & 2025-04-12 & 400-M1 & 1'' & 3800-7000 \AA & 3 $\times$ 300\,s & 10.6 & DA\\
WD\,J205208.50$-$501335.57 & 2025-04-12 & 400-M1 & 1'' & 3800-7000 \AA & 4 $\times$ 480\,s & 7.6 & DH\\
WD\,J212401.83$-$600059.54 & 2025-04-23 & 400-M1 & 1'' & 3800-7000 \AA & 4 $\times$ 480\,s & 11.7 & DA\\
WD\,J215201.03$-$724022.50 & 2025-03-22 & 400-M1 & 1'' & 3800-7000 \AA & 4 $\times$ 300\,s & 9.5 & DA\\
WD\,J230658.73$-$290514.86 & 2025-06-05 & 400-M1 & 1'' & 3800-7000 \AA & 4 $\times$ 300\,s & 9.7 & DA\\
\enddata
\tablecomments{All SOAR/Goodman 400-M1 use the Red camera unless noted $^*$ if using the Blue camera. We derive S/N ratios by dividing the mean spectrum flux by the spectrum standard deviation in a 200 \AA\, window around 4600 \AA.}
\end{deluxetable*}

\onecolumngrid
\section{Photometry}

\renewcommand{\thefigure}{B.\arabic{figure}}
\setcounter{figure}{0}

\begin{figure*}[t]
  \centering
  {\includegraphics[width=0.95\textwidth]{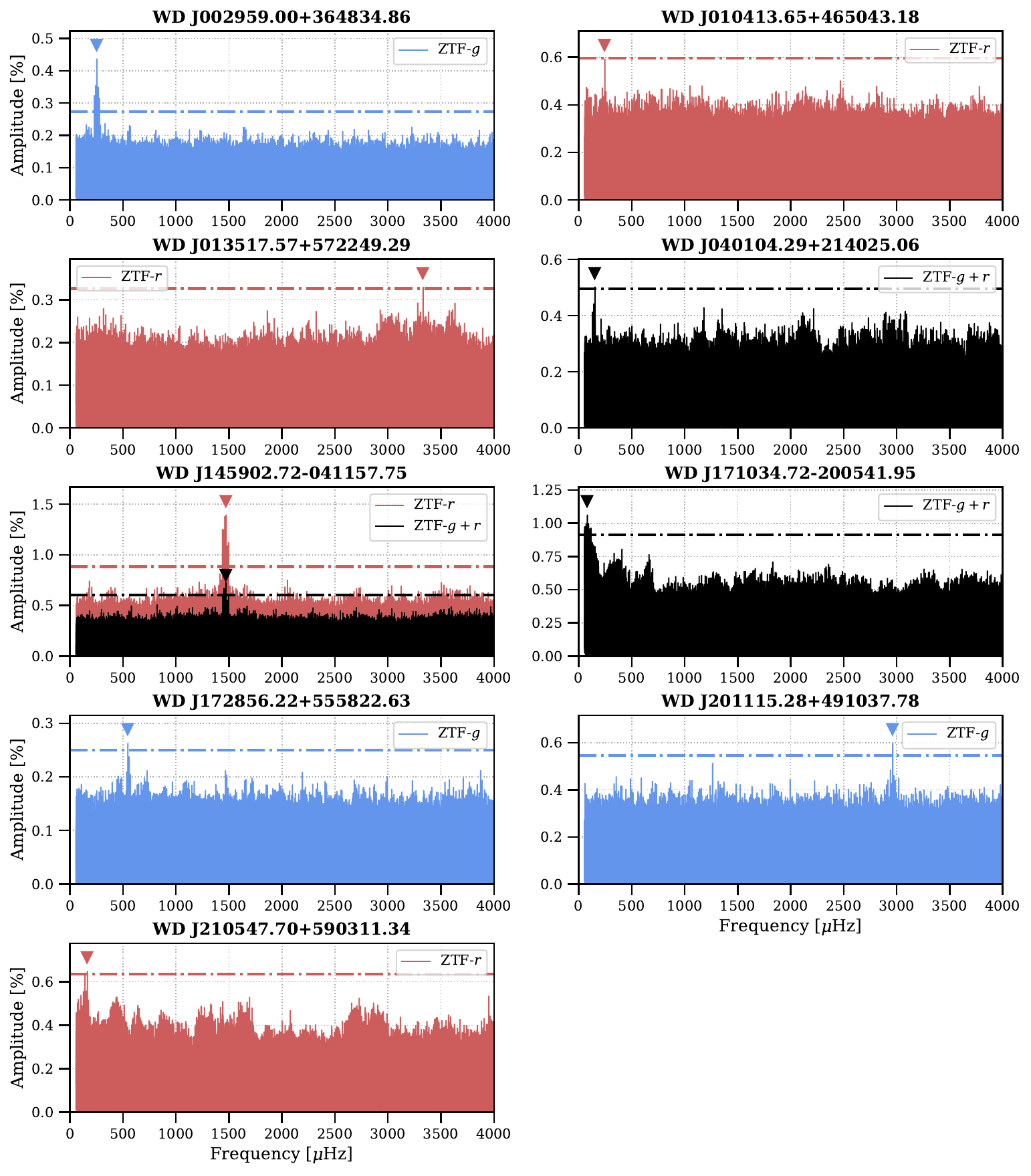}}
  \caption{Lomb-Scargle periodograms of all the objects that show significant periodic variability in their ZTF light curves, as detailed in Table~\ref{tab:ztf_phot}. The dashed-dotted lines are the 0.1\% false-alarm probability significance thresholds we bootstrapped.} \label{fig:ztf_lsps}
\end{figure*}

Here we summarize all our follow-up time series photometry observations and take account of the objects that show significant periodic variability in their ZTF light curves. Table~\ref{tab:ztf_phot} includes the detected periods for 10 objects showing significant variability in at least one of the ZTF combined $g$, $r$, or $g+r$, excluding objects with significant variability at the harmonics of 1\,d due to aliasing. We plot the corresponding Lomb-Scargle periodograms of these systems in Figure~\ref{fig:ztf_lsps}.

\begin{figure*}[t]
  \centering
  {\includegraphics[width=0.95\textwidth]{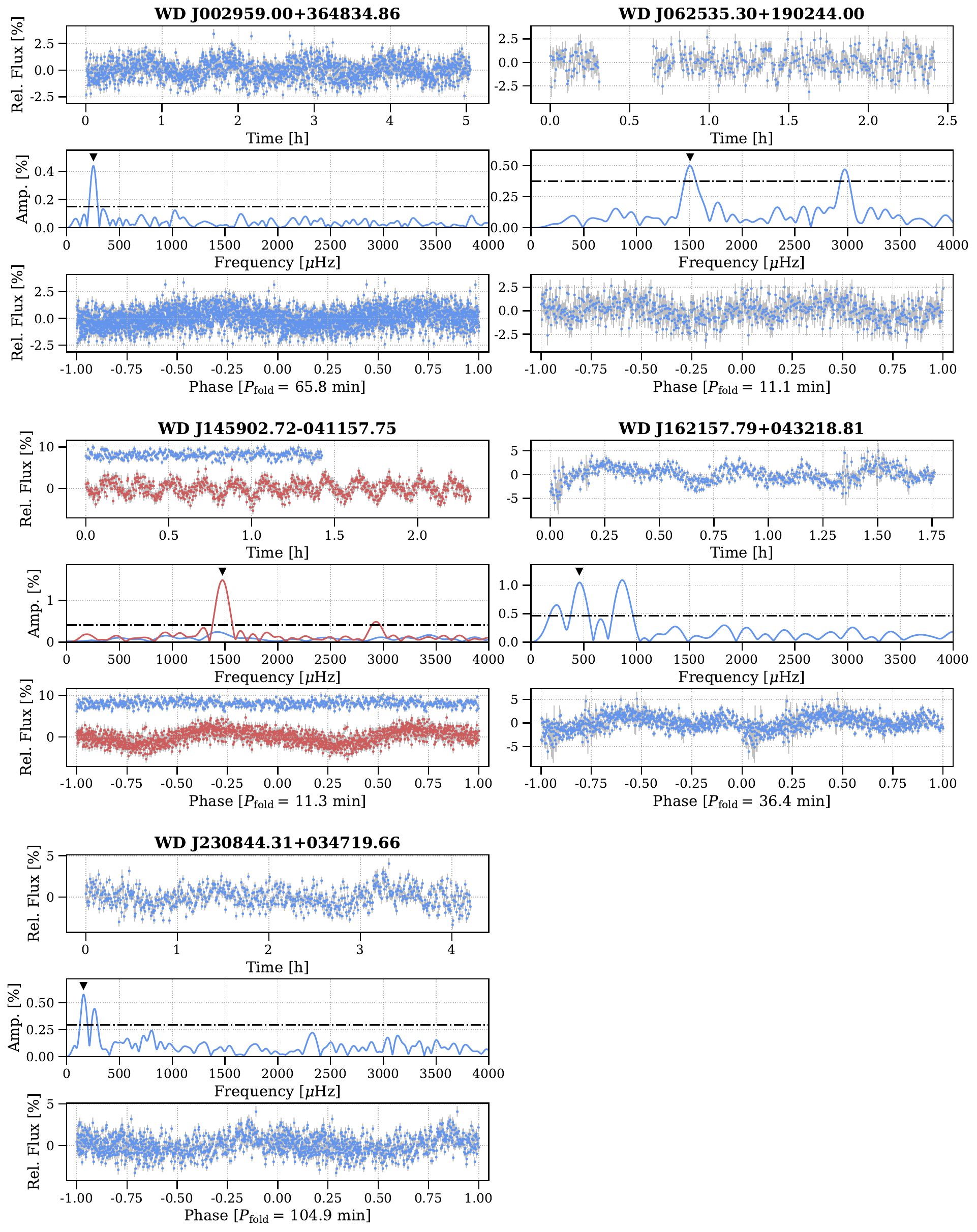}}
  \caption{Light curves and periodograms of all five rapid rotator candidates from the young subset of the Q branch region as described in Section~\ref{sec:variability}. We also fold the light curves on the most dominant signal returned, marked by black triangles on the periodograms. All photometry plotted in blue were collected using a {\em BG40} filter, while the red were taken with the SDSS-$r$ filter. The black dashed-dotted lines are the 0.1\% false-alarm probability significance thresholds we bootstrapped.} \label{fig:LC}
\end{figure*}

\begin{figure*}[t]
  \centering
  {\includegraphics[width=0.95\textwidth]{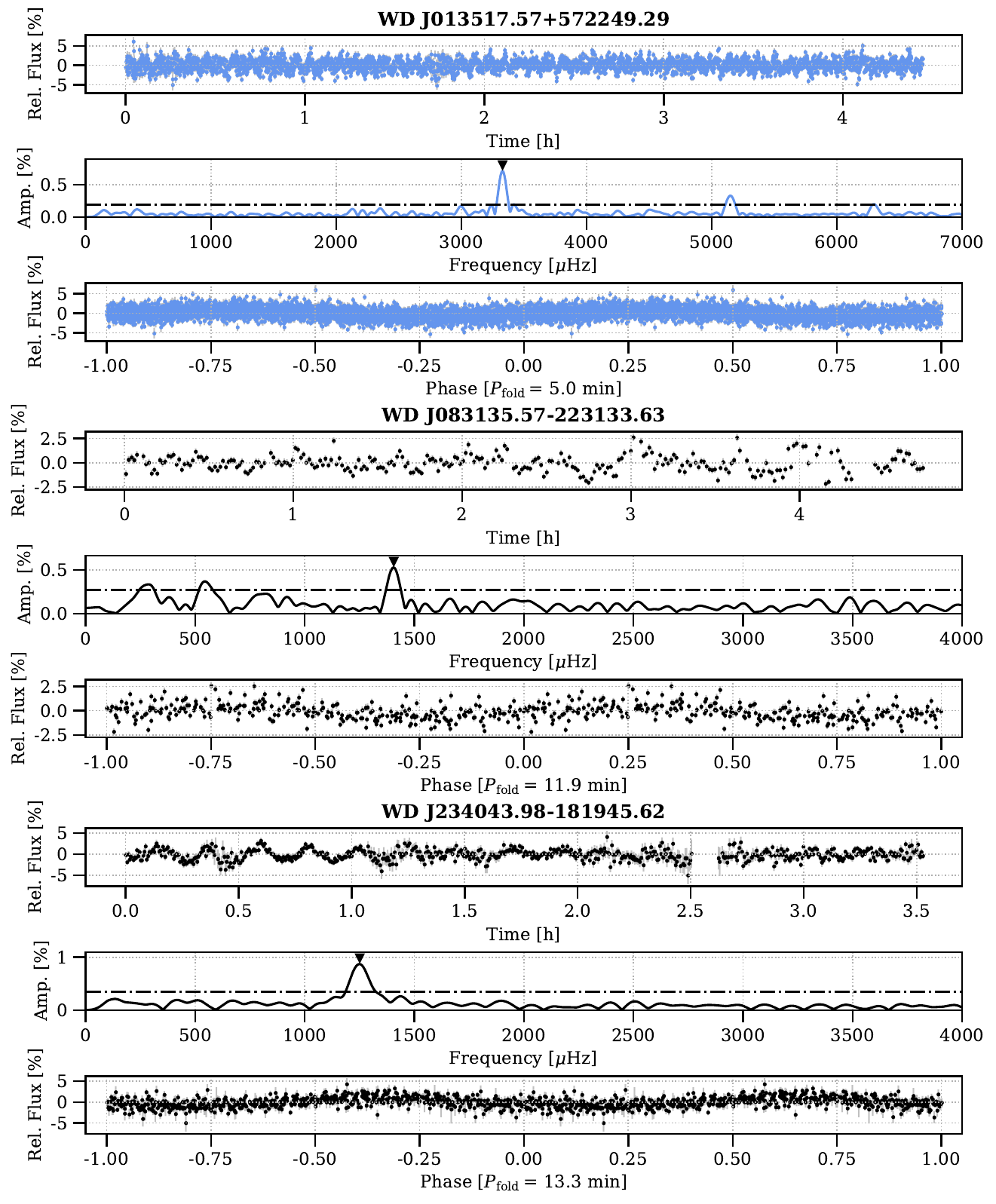}}
  \caption{Light curves and Lomb-Scargle periodograms of the three pulsating Q branch objects in our sample that we observed, including the newly discovered pulsating DAQs WD\,J234043.98$-$181945.62 and WD\,J083135.57$-$223133.63, all as detailed in Section~\ref{sec:pulsating}. In the same fashion as Figure~\ref{fig:LC}, we also fold the light curves on the most dominant signal returned, marked by black triangles on the periodograms. All photometry plotted in blue were collected using a {\em BG40} filter, while the black were taken with the SDSS-$g$ filter. The black dashed-dotted lines are the 0.1\% false-alarm probability significance thresholds we bootstrapped. The photometry for WD\,J083135.57$-$223133.63 is binned by a factor of three to break the 1:36.5 resonance between the effective exposure time (19.53\,s) and the period, which only allowed us to sample the same 73 discrete phase bins.}\label{fig:LC_pulsators}
\end{figure*}

Figure~\ref{fig:LC} shows the light curves and associated Lomb-Scargle periodograms for all five rapid rotator in the young Q branch sample, with Figure~\ref{fig:LC_pulsators} as is its analog for the two pulsating objects in our sample. The observational details of each object is described in Table~\ref{tab:photometry}, where we also share a record of the entirety of our follow-up campaign.

\begin{deluxetable*}{lcccc}[htbp!]
\tablenum{B1}
\tablecolumns{5}
\tablecaption{ZTF period search \label{tab:ztf_phot}}
\tabletypesize{\footnotesize}
\tablehead{
    \colhead{Target}    & \colhead{$N_{\mathrm{obs}}$} &
    \colhead{$P$} & \colhead{$A$} &  \colhead{Significance Threshold} \\ [-0.2cm]
    \colhead{} & \colhead{} & \colhead{s} & \colhead{\%} & \colhead{\%}
}
\startdata
\cutinheadnew{$g$-band Significant Periods}
WD\,J002959.00+364834.86 & 2060 & 3983.9 & 0.42 & 0.27\\
WD\,J172856.22+555822.63 & 1739 & 1838.4 & 0.26 & 0.25\\
WD\,J201115.28+491037.78 & 665 & 337.8 & 0.60 & 0.54\\
\hline
\cutinheadnew{$r$-band Significant Periods}
WD\,J010413.65+465043.18 & 1101 & 4113.6 & 0.60 & 0.60\\
WD\,J013517.57+572249.29 & 1852 & 300.5 & 0.33 & 0.33\\
WD\,J145902.72$-$041157.75 & 383 & 680.1 & 1.39 & 0.88\\
WD\,J210547.70+590311.34 & 1177 & 6230.7 & 0.65 & 0.64\\
\hline
\cutinheadnew{$g+r$ Significant Periods}
WD\,J040104.29+214025.06 & 1130 & 6650.0 & 0.50 & 0.50\\
WD\,J145902.72$-$041157.75 & 653 & 680.1 & 0.73 & 0.60\\
WD\,J171034.72$-$200541.95 & 895 & 12758.2 & 1.06 & 0.91\\
\enddata
\end{deluxetable*}

\begin{longrotatetable}
\begin{deluxetable*}{lccccccccccc}
\tablenum{B2}
\tablecolumns{11}
\tablecaption{Observing Log of Follow-up Time-Series Photometry Observations \label{tab:photometry}}
\tabletypesize{\footnotesize}
\tablehead{
    \colhead{Target}    & \colhead{UT Date}  & \colhead{Facility} &
    \colhead{Duration}   & \colhead{$t_{\mathrm{exp}}$} & 
    \colhead{Filter}     & \colhead{Seeing} & 
    \colhead{$f$} & \colhead{$A$} &  \colhead{S.T.} & \colhead{Variability}  \\ [-0.2cm]
    \colhead{}           & \colhead{}      & \colhead{} &
    \colhead{hr}       & \colhead{s}   &  \colhead{} &
    \colhead{arcsec}   & \colhead{$\mu$Hz}    &  \colhead{\%} &  \colhead{\%} & \colhead{}
}
\startdata
\cutinheadnew{Delayed Subset}
WD\,J004527.50$-$233629.29 & 2024-12-21 & PTO/PRISM & 2.30 & 25 & {\em BG40} & 3.5 & - & - & 0.93 & \\ 
WD\,J010413.65+465043.18 & 2024-11-24 & McD/ProEM & 4.78 & 15 & {\em BG40} & 1.1 & - & - & 0.20 & \\ 
WD\,J020549.70+205707.96 & 2024-11-26 & McD/ProEM & 1.75 & 20 & {\em BG40} & 1.9 & - & - & 0.63 & \\ 
WD\,J052950.25+523953.39 & 2024-11-26 & McD/ProEM & 3.40 & 10 & {\em BG40} & 1.6 & - & - & 0.22 & \\ 
WD\,J065535.29+293909.66 & 2025-01-29 & McD/ProEM & 1.28 & 10 & {\em BG40} & 2.1 & - & - & 0.48 & \\ 
WD\,J080122.81+774957.21 & 2024-11-25 & McD/ProEM & 1.40 & 5 & {\em BG40} & 1.7 & - & - & 0.72 & \\
WD\,J083135.57$-$223133.63 & 2026-01-17 & LDT/LMI & 4.74 & 15 &  $g$ & 3.17 & 1402.7 $\pm$ 3.5 & 0.54 $\pm$ 0.06 & 0.28 & Pulsations\\
WD\,J170618.72$-$083747.26 & 2025-09-18 & LDT/LMI & 1.43 & 15 & {\em g} & 1.7 & - & - & 0.62 & \\ 
WD\,J172856.22+555822.63 & 2025-06-28 & McD/ProEM & 2.56 & 20 & {\em BG40} & 1.0 & - & - & 0.70 & \\ 
WD\,J234043.98$-$181945.62 & 2025-09-18 & LDT/LMI & 3.52 & 15 & {\em g} & 1.3 & 1252.3 $\pm$ 2.8 & 1.16 $\pm$ 0.07 & 0.35 & Pulsations\\
 & & & & & & & 1275.5 $\pm$ 4.4 & 0.73 $\pm$ 0.07 & & \\
\hline
\cutinheadnew{Young Subset}
WD\,J002959.00+364834.86 & 2024-11-23 & McD/ProEM & 5.04 & 10 & {\em BG40} & 1.0 & 252.4 $\pm$ 2.1 & 0.44 $\pm$ 0.03 & 0.15 & Rapid Rotation\\ 
WD\,J010745.24+290420.82 & 2025-01-02 & McD/ProEM & 2.18 & 10 & {\em BG40} & 1.2 & - & - & 0.20 & \\ 
WD\,J013517.57+572249.29 & 2024-11-25 & McD/ProEM & 4.44 & 5 & {\em BG40} & 1.1 & 3329.1 $\pm$ 1.8 & 0.71 $\pm$ 0.04 & 0.19 & Pulsations\\
 & & & & & & & 7257.2 $\pm$ 3.4 & 0.38 $\pm$ 0.04 & & \\
 & & & & & & & 5147.0 $\pm$ 3.8 & 0.34 $\pm$ 0.04 & & \\
 & & & & & & & 6297.7 $\pm$ 6.5 & 0.20 $\pm$ 0.04 & & \\
WD\,J025001.75$-$043703.17 & 2025-01-03 & McD/ProEM & 1.69 & 15 & {\em BG40} & 1.4 & - & - & 0.48 & \\ 
WD\,J062535.30+190244.00 & 2024-11-23 & McD/ProEM & 2.41 & 20 & {\em BG40} & 1.7 & 1506.0 $\pm$ 10.8 & 0.49 $\pm$ 0.08 & 0.86 & Rapid Rotation\\ 
  & & & & & & & 2976.6 $\pm$ 11.7 & 0.46 $\pm$ 0.08 & & \\
WD\,J095057.58$-$284115.39 & 2025-05-29 & McD/ProEM & 0.79 & 15 & {\em BG40} & 2.7 & - & - & 1.21 & \\ 
WD\,J113502.24$-$243020.09 & 2025-03-26 & McD/ProEM & 1.15 & 30 & {\em BG40} & 1.5 & - & - & 7.79 & \\ 
WD\,J133340.35+640627.35 & 2025-07-25 & McD/ProEM & 2.62 & 20 & {\em BG40} & 1.8 & - & - & 0.88 & \\ 
WD\,J145902.72$-$041157.75 & 2025-01-30 & McD/ProEM & 1.42 & 15 & {\em BG40} & 3.4 & - & - & 0.29 & \\ 
 & 2025-01-31 & McD/ProEM & 2.31 & 10 & {\em r} & 1.8 & 1475.3 $\pm$ 3.8 & 1.51 $\pm$ 0.09 & 0.41 & Rapid Rotation\\ 
   & & & & & & & 2932.3 $\pm$ 10.9 & 0.53 $\pm$ 0.09 & &  \\
WD\,J162157.79+043218.81 & 2025-05-29 & McD/ProEM & 1.76 & 10 & {\em BG40} & 1.4 & 869.2 $\pm$ 8.4 & 1.01 $\pm$ 0.10 & 0.46 & Rapid Rotation\\
  & & & & & & & 465.6 $\pm$ 10.2 & 0.83 $\pm$ 0.10 & & \\
  & & & & & & & 190.9 $\pm$ 16.6 & 0.51 $\pm$ 0.10 & & \\
WD\,J184926.21+645810.98 & 2025-06-30 & McD/ProEM & 1.52 & 20 & {\em BG40} & 1.9 & - & - & 0.71 & \\ 
WD\,J192555.20$-$034626.55 & 2025-10-24 & PTO/PRISM & 1.91 & 20 & {\em BG40} & 2.8 & - & - & 0.32 & \\
WD\,J215331.59$-$262855.53 & 2025-10-24 & PTO/PRISM & 1.52 & 20 & {\em BG40} & 3.5 & - & - & 0.66 & \\ 
WD\,J230844.31+034719.66 & 2024-11-27 & McD/ProEM & 3.38 & 20 & {\em BG40} & 1.4 & 154.6 $\pm$ 4.6 & 0.50 $\pm$ 0.06 & 0.30 & Rapid Rotation\\
  & & & & & & & 271.3 $\pm$ 6.7 & 0.35 $\pm$ 0.06 & & \\
WD\,J234921.84+090714.70 & 2025-09-17 & PTO/PRISM & 2.05 & 15 & {\em BG40} & 3.0 & - & - & 0.57 & \\
\enddata
\tablecomments{We derive average seeing estimates from the FWHM found at each image by {\tt hipercam} during our extraction. The frequencies and amplitudes we report are obtained from LLS fits to the time series using {\tt pyriod}. For objects with multiple periodicities, we list them in order by amplitude. We estimate uncertainties on these parameters using the framework provided by \citet{1999DSSN...13...28M}. S.T. is our abbreviation for the 0.1\% false-alarm probability significance thresholds we bootstrap.}
\end{deluxetable*}
\end{longrotatetable}

\onecolumngrid
\section{Q branch Sample}

We present our full Q branch sample within 100\,pc in Table~\ref{tab:sample}, highlighting both the delayed subset and the young subset. We include the target names and Gaia source id, the publication source for the target spectrum, the distance (pc) reported in \citet{2018AJ....156...58B}, the transverse velocity $v_{\rm t}$ as calculated from Equation~\ref{eq}. We report the bootstrapped NOV variability limits from our photometric follow-up as detailed in Table~\ref{tab:photometry} when available. For the remaining targets, we report the  bootstrapped NOV variability limits from ZTF for targets at declinations higher than 30 deg, or ATLAS for targets at declinations lower than 30 deg. Finally, we include the spectral type of each target. All corresponding uncertainties are included. The targets are ranked alphabetically in each subset.

\renewcommand{\thetable}{C\arabic{table}}

\begin{longrotatetable}
\begin{deluxetable*}{lccccccc}
\tablenum{C1}
\tablecolumns{7}
\tablecaption{Full Q branch sample within 100\,pc \label{tab:sample}}
\tabletypesize{\footnotesize}
\tablehead{
    \colhead{Target}    & \colhead{Gaia source id}  & \colhead{Spec. source} &
    \colhead{Distance}   & \colhead{$v_{\rm t}$} & 
    \colhead{Sig. Thresh}     & \colhead{Spectral Type} \\ [-0.2cm]
    \colhead{}             & \colhead{}             & \colhead{}             & \colhead{pc}           & 
    \colhead{km\,s$^{-1}$} & \colhead{\%}           & \colhead{} 
}
\startdata
\cutinheadnew{\textbf{Delayed Subset}}
WD\,J004527.50-233629.29 & 2348747743931814656 & This Work & 47.4 (0.3) & 71.3 (0.3) & 0.91 & DQ \\
WD\,J010413.65+465043.18 & 401215160231429120 & \citet{2020ApJ...898...84K} & 83.8 (1.1) & 50.3 (0.5) & 0.2 & DQ \\
WD\,J012709.06-575914.45 & 4909586088245414912 & This Work & 93.8 (0.9) & 72.3 (0.5) & 3.44 & DA \\
WD\,J020549.70+205707.96 & 94276941624384000 & \citet{2024ApJ...974...12J} & 85.4 (1.0) & 133.1 (1.3) & 0.63 & DAQ \\
WD\,J034703.18-180253.49 & 5107322396824711680 & \citet{2024ApJ...974...12J} & 76.4 (0.8) & 58.9 (0.5) & 1.35 & DA \\
WD\,J040104.29+214025.06 & 53042472446551424 & \citet{2024ApJ...974...12J} & 71.4 (0.6) & 88.4 (0.5) & 0.5 & DA \\
WD\,J044747.69+422438.69 & 203678825329613312 & \citet{2024ApJ...974...12J} & 95.7 (1.6) & 88.5 (1.0) & 1.78 & DA \\
WD\,J045533.97-005815.35 & 3226519762223501696 & This Work & 91.7 (1.3) & 91.9 (1.0) & 3.48 & DA \\
WD\,J052724.33-310655.56 & 2905067333001582720 & \citet{2007AJ....134..252S} & 40.9 (0.1) & 69.2 (0.1) & 1.21 & DA \\
WD\,J052950.25+523953.39 & 263082591016645504 & \citet{2011ApJ...743..138G} & 38.9 (0.1) & 122.1 (0.2) & 0.2 & DA \\
WD\,J065535.29+293909.66 & 887758130788405504 & \citet{2024ApJ...974...12J} & 79.4 (0.7) & 71.9 (0.5) & 0.41 & DAQ \\
WD\,J080122.81+774957.21 & 1138353770109251200 & This Work & 79.4 (0.6) & 55.2 (0.3) & 0.4 & DA \\
WD\,J083135.57-223133.63 & 5702793425999272576 & \citet{2020ApJ...898...84K} & 82.0 (0.9) & 99.1 (0.7) &  & DAQ \\
WD\,J110703.79+040543.60 & 3815200997858084480 & \citet{2020ApJ...898...84K} & 62.4 (0.5) & 68.7 (0.4) & 1.58 & DA \\
WD\,J120331.90+645101.41 & 1585063422960992256 & SDSS & 87.1 (0.8) & 63.3 (0.5) & 0.5 & DQA \\
WD\,J134347.63-442902.53 & 6108445596683010688 & This Work & 97.8 (2.4) & 112.7 (1.5) & 12.97 & DQA \\
WD\,J151949.65+632951.86 & 1643814211883928320 & \citet{2011ApJ...743..138G} & 55.7 (0.2) & 61.7 (0.1) & 0.23 & DA \\
WD\,J161556.15-395346.09 & 5994123531590244480 & This Work & 75.3 (0.7) & 77.9 (0.6) & 3.46 & DA \\
WD\,J162236.25+300455.29 & 1318204460477280512 & SDSS & 74.7 (0.5) & 53.1 (0.3) & 0.25 & DQA \\
WD\,J165915.38+661032.66 & 1635687790163070976 & \citet{2011ApJ...743..138G} & 64.5 (0.3) & 59.2 (0.2) &  & DA \\
WD\,J170145.15-524609.22 & 5935933154350863872 & This Work & 40.5 (0.1) & 59.7 (0.1) & 3.96 & DAQ \\
WD\,J170618.72-083747.26 & 4336571785203401472 & \citet{2024ApJ...974...12J} & 67.7 (0.8) & 136.7 (1.0) & 0.76 & DQA \\
WD\,J171034.72-200541.95 & 4128167950420485632 & \citet{2024ApJ...974...12J} & 87.3 (1.8) & 52.6 (0.6) & 0.91 & DQA \\
WD\,J172856.22+555822.63 & 1422012892308493568 & SDSS & 47.1 (0.1) & 56.3 (0.1) & 0.18 & DQA \\
WD\,J183302.73-412058.38 & 6723335899591792512 & This Work & 74.8 (0.6) & 58.2 (0.4) & 3.78 & DA \\
WD\,J184809.11-461958.62 & 6705552153700215936 & This Work & 93.9 (1.3) & 98.4 (1.0) & 5.92 & DA \\
WD\,J192407.28-271750.43 & 6765861019327924736 & \citet{2024ApJ...974...12J} & 70.7 (0.8) & 84.3 (0.7) & 0.54 & DA \\
WD\,J201115.28+491037.78 & 2087569060381096960 & \citet{2024ApJ...974...12J} & 73.6 (0.5) & 58.0 (0.3) & 0.34 & DQ \\
WD\,J214023.96-363757.44 & 6589369272547881856 & \citet{2007AJ....134..252S} & 39.7 (0.2) & 55.5 (0.1) & 1.62 & DQ \\
WD\,J234043.98-181945.62 & 2393834386459511680 & \citet{2020ApJ...898...84K} & 96.1 (2.1) & 51.1 (0.7) &  & DAQ \\
\\
\hline
\cutinheadnew{\textbf{Young Subset}}
WD\,J001534.13-604927.14 & 4904968208127215104 & This Work & 73.9 (0.6) & 37.9 (0.2) & 3.61 & DH \\
WD\,J002959.00+364834.86 & 366784816895496064 & \citet{2020ApJ...898...84K} & 59.5 (0.6) & 28.4 (0.1) &  & DA \\
WD\,J010745.24+290420.82 & 308383019835259008 & \citet{2020ApJ...898...84K} & 59.1 (0.4) & 14.1 (0.1) & 0.19 & DA \\
WD\,J013517.57+572249.29 & 412839403319209600 & \citet{2020ApJ...898...84K} & 51.0 (0.2) & 29.1 (0.1) &  & DA \\
WD\,J025001.75-043703.17 & 5184589747536175104 & SDSS & 91.3 (1.5) & 10.7 (0.1) & 0.52 & DH \\
WD\,J032547.53-081549.36 & 5168129306849447552 & \citet{2024ApJ...974...12J} & 86.5 (1.2) & 11.5 (0.1) & 0.84 & DA \\
WD\,J042226.08-240724.12 & 4896474240286811648 & \citet{2024ApJ...974...12J} & 71.6 (0.4) & 29.4 (0.1) & 1.9 & DA \\
WD\,J045919.46-540320.77 & 4782830054172758272 & This Work & 92.2 (1.0) & 38.6 (0.3) & 3.08 & DA \\
WD\,J050709.66+264515.13 & 3421894079307215744 & \citet{2015MNRAS.454.2787G} & 53.8 (0.2) & 11.7 (0.0) & 0.42 & DAH \\
WD\,J055134.45+413529.95 & 192275966334956672 & \citet{2020ApJ...898...84K} & 46.4 (0.2) & 29.8 (0.1) &  & DAQ \\
WD\,J062535.30+190244.00 & 3372355922219793536 & \citet{2024ApJ...974...12J} & 94.9 (1.6) & 17.7 (0.3) &  & DH \\
WD\,J065711.00+734144.97 & 1114813977776610944 & \citet{2024ApJ...974...12J} & 82.9 (1.0) & 15.2 (0.1) & 0.67 & DA \\
WD\,J072026.97-413924.27 & 5560637430205458304 & This Work & 97.3 (1.3) & 32.3 (0.3) & 10.68 & DH \\
WD\,J073744.49-061018.89 & 3055999768051212416 & \citet{2024ApJ...974...12J} & 77.4 (0.9) & 12.6 (0.1) & 0.55 & DQ \\
WD\,J081149.35+421208.95 & 921804126089222784 & \citet{2011ApJ...743..138G} & 91.8 (1.2) & 36.1 (0.4) & 0.55 & DA \\
WD\,J094500.96-685304.67 & 5243591401210032000 & This Work & 85.5 (0.9) & 16.7 (0.1) & 3.81 & DQ \\
WD\,J094928.05-514958.10 & 5405554510567061120 & This Work & 91.5 (1.0) & 28.7 (0.2) & 8.51 & DA \\
WD\,J095057.58-284115.39 & 5464929134894103808 & \citet{2024ApJ...974...12J} & 78.8 (0.7) & 12.9 (0.1) & 1.33 & DA \\
WD\,J095328.98-671447.92 & 5245657246123148672 & This Work & 92.7 (1.2) & 17.5 (0.2) & 4.7 & DH \\
WD\,J100219.81-371352.08 & 5422144977092149632 & This Work & 43.2 (0.1) & 26.4 (0.0) & 2.15 & DH \\
WD\,J110752.92-160705.38 & 3559496413333223936 & SDSS & 83.9 (1.3) & 48.8 (0.5) & 2.33 & DA \\
WD\,J113502.24-243020.09 & 3533885454629567872 & \citet{2024ApJ...974...12J} & 75.7 (0.9) & 41.2 (0.3) & 2.1 & DA \\
WD\,J115618.17-675029.94 & 5235637774642886272 & This Work & 90.7 (1.0) & 43.8 (0.3) &  & DA \\
WD\,J133340.35+640627.35 & 1665858350572796672 & SDSS & 100.0 (1.4) & 31.9 (0.3) & 0.63 & DAH \\
WD\,J145902.72-041157.75 & 6338900661178928896 & \citet{2024ApJ...974...12J} & 60.7 (0.6) & 22.9 (0.1) &  & DH \\
WD\,J162157.79+043218.81 & 4436905352274528896 & \citet{2020ApJ...898...84K} & 61.3 (0.3) & 19.7 (0.1) &  & DH \\
WD\,J162659.59+253326.79 & 1304081783374935680 & \citet{2020ApJ...898...84K} & 83.9 (0.7) & 23.8 (0.1) &  & DA \\
WD\,J175253.90-663452.57 & 5812676526434412928 & This Work & 53.8 (0.2) & 45.9 (0.1) & 3.17 & DQ \\
WD\,J175931.34-620108.87 & 5911160263973498496 & \citet{2023MNRAS.518.3055O} & 38.3 (0.1) & 20.0 (0.0) & 3.88 & DA \\
WD\,J183449.52-450730.01 & 6709422675147738752 & This Work & 89.0 (1.3) & 12.0 (0.1) & 5.5 & DA \\
WD\,J184926.21+645810.98 & 2253826832091026560 & \citet{2020ApJ...898...84K} & 86.4 (0.7) & 12.9 (0.1) & 0.32 & DH \\
WD\,J192555.20-034626.55 & 4213471120498390784 & \citet{2024ApJ...974...12J} & 56.8 (0.3) & 27.2 (0.1) & 0.34 & DQA \\
WD\,J193431.84-653025.80 & 6440255975194724736 & This Work & 96.9 (1.6) & 43.9 (0.5) & 4.41 & DQA \\
WD\,J194736.29-310038.84 & 6751474223204472576 & This Work & 72.4 (0.9) & 23.6 (0.2) & 1.92 & DAH \\
WD\,J204453.39-432919.16 & 6676948053760503296 & This Work & 49.9 (0.2) & 14.2 (0.0) & 1.7 & DA \\
WD\,J205208.50-501335.57 & 6477825157941375232 & This Work & 100.0 (1.6) & 36.7 (0.5) & 4.7 & DH \\
WD\,J205442.80-203925.93 & 6857295585945072128 & \citet{2011ApJ...743..138G} & 31.2 (0.0) & 10.3 (0.0) & 0.33 & DA \\
WD\,J210547.70+590311.34 & 2190645256129430144 & \citet{2024ApJ...974...12J} & 95.7 (1.0) & 26.4 (0.2) & 0.53 & DA \\
WD\,J212401.83-600059.54 & 6452369226077816064 & This Work & 90.4 (1.5) & 44.6 (0.4) & 5.83 & DA \\
WD\,J215201.03-724022.50 & 6371070309824311808 & This Work & 92.2 (0.9) & 12.7 (0.1) & 3.92 & DA \\
WD\,J215331.59-262855.53 & 6811977801160882944 & \citet{2024ApJ...974...12J} & 79.2 (1.0) & 16.3 (0.1) & 0.73 & DA \\
WD\,J222609.28-311141.35 & 6602096222717844224 & \citet{2004MNRAS.349.1397C} & 95.4 (1.4) & 23.3 (0.3) & 2.77 & DA \\
WD\,J230658.73-290514.86 & 6606362529696816896 & This Work & 95.2 (1.9) & 5.5 (0.1) & 2.47 & DA \\
WD\,J230844.31+034719.66 & 2662208372887759744 & \citet{2020ApJ...898...84K} & 64.6 (0.6) & 28.6 (0.2) &  & DQA \\
WD\,J234921.84+090714.70 & 2758938385082121216 & \citet{2020ApJ...898...84K} & 86.1 (1.1) & 14.1 (0.1) & 0.63 & DA \\
\enddata
\tablecomments{Distances and confidence interval of the estimated distance are reported from \citet{2018AJ....156...58B}. Sig. Thresh is the abbreviation for the 0.1\% false-alarm probability significance thresholds we bootstrap. The values for targets at declinations lower than 30 deg are from ATLAS. Otherwise, we pick the lower value between ZTF and our follow-up described in Table~\ref{tab:photometry}.}
\end{deluxetable*}
\end{longrotatetable}


\clearpage

\bibliography{bibliography}{}
\bibliographystyle{aasjournal}

\end{CJK*}
\end{document}